\documentclass[reprint, amsmath, amssymb, aps, showkeys]{revtex4-2}
\usepackage{graphicx}
\usepackage{dcolumn}
\usepackage{bm}
\usepackage{amsmath}
\usepackage{float}
\usepackage{subfigure}
\usepackage{booktabs}
\usepackage{CJKutf8}
\usepackage{hyperref}
\usepackage{tikz,xcolor}

\hypersetup{
  colorlinks=true,
  linkcolor=blue,
  urlcolor=blue,
  citecolor=blue,
  linktoc=all
}
\definecolor{lime}{HTML}{A6CE39}
\DeclareRobustCommand{\orcidicon}{%
	\begin{tikzpicture}
	\draw[lime, fill=lime] (0,0) 
	circle [radius=0.16] 
	node[white] {{\fontfamily{qag}\selectfont \tiny ID}};
	\draw[white, fill=white] (-0.0625,0.095) 
	circle [radius=0.007];
	\end{tikzpicture}
	\hspace{-2mm}
}

\foreach \x in {A, ..., Z}{%
	\expandafter\xdef\csname orcid\x\endcsname{\noexpand\href{https://orcid.org/\csname orcidauthor\x\endcsname}{\noexpand\orcidicon}}
}

\begin{document}
\begin{CJK*}{UTF8}{gbsn}

\title{Comparison and application of different post-Newtonian models for inspiralling stellar-mass binary black holes with space-based GW detectors}

\author{Jie Wu (吴洁)\orcidA{}$^{1,2}$ }
\author{Jin Li (李瑾)\orcidB{}$^{1,2}$ }
\email{cqujinli1983@cqu.edu.cn}
\author{Xiaolin Liu (刘骁麟)\orcidC{}$^{3}$ }
\author{Zhoujian Cao (曹周键)\orcidD{}$^{4,5,6}$ }

\affiliation{$^{1}$College of Physics, Chongqing University, Chongqing 401331, China}
\affiliation{$^{2}$Department of Physics and Chongqing Key Laboratory for Strongly Coupled Physics, Chongqing University, Chongqing 401331, China}
\affiliation{$^{3}$Institudo de Física Téorica UAM-CSIC, Universidad Autónoma de Madrid, 28049, Spain}
\affiliation{$^{4}$Department of Astronomy, Beijing Normal University, Beijing 100875, China}
\affiliation{$^{5}$Institute for Frontiers in Astronomy and Astrophysics, Beijing Normal University, Beijing 102206, China}
\affiliation{$^{6}$School of Fundamental Physics and Mathematical Sciences, Hangzhou Institute for Advanced Study, UCAS, Hangzhou 310024, China}

\begin{abstract}
Space-based gravitational wave (GW) detectors are expected to detect the stellar-mass binary black holes (SBBHs) inspiralling in the low-frequency band, which exist in several years before the merger.
Accurate GW waveforms in the inspiral phase are crucial for the detection and analysis of those SBBHs.
In our study, based on post-Newtonian (PN) models, we investigate the differences in the detection, accuracy requirement, and parameter estimation of SBBHs in the cases of LISA, Taiji, and their joint detection.
We find that low-order PN models are sufficient for simulating low-mass ($\le 50\ \mathrm{M}_\odot$) SBBHs population.
Moreover, for detectable SBBHs in space-based GW detectors, over 90\% of the GW signals from low-order PN models meet accuracy requirement. 
Additionally, different PN models do not exhibit significant differences in Bayesian inference.
Our research provides a comprehensive reference for balancing computational resources and the desired accuracy of GW waveform generation.
It highlights the suitability of low-order PN models for simulating SBBHs and emphasizes their potential in the detection and parameter estimation of SBBHs.
\end{abstract}

\maketitle
\end{CJK*}

\section{Introduction}
Gravitational waves (GWs) have revolutionized our understanding of the Universe, providing us with a new observation window.
Among the various sources, one of the most significant GW sources are stellar-mass binary black holes (SBBHs), also referred to as stellar-origin binary black holes, typically with primary component masses ranging from a few~$\mathrm{M_{\odot }}$ up to $\sim100\ \mathrm{M_{\odot }}$.
In 2015, the Advanced LIGO first detected a GW event generated by the merger of a SBBH with component masses 35.6~$\mathrm{M_{\odot }}$ and 30.6~$\mathrm{M_{\odot }}$~\cite{GW150914}.
In the following years, GW events from binary neutron star~\cite{BNS1,BNS2} and neutron star-black hole~\cite{NSBH} were successively detected, opening the new area of multimessenger astronomy~\cite{Multi-messenger1,Multi-messenger2,Multi-messenger3,Multi-messenger4}.
As of this year, the LIGO-Virgo-KAGRA (LVK) collaboration has detected over 90 GW events, which are all from stellar-mass compact binary systems~\cite{GWTC1,GWTC2,GWTC3}, and commenced the Fourth Observing Run (\href{https://observing.docs.ligo.org/plan/}{O4}).

The GWs generated by SBBHs span multiple frequency bands, encompassing the inspiral phase in the low-frequency bands and the merger phase in the high-frequency bands~\cite{network}.
Ground-based GW detectors like LVK have the potential to detect the GWs emitted in the merger phase (or more precisely, the very late inspiral, merger, and ringdown phases) of SBBHs, which can last for several seconds to several minutes at most~\cite{aLIGO,aVIRGO,KAGRA}.
Before the merger, a significant number of SBBHs are inspiralling in the low-frequency band, with the expectation of entering the sensitivity range of space-based GW detectors.
Space-based GW detectors consist of various missions, such as the LISA mission proposed by the European Space Agency and the Taiji mission proposed by the Chinese Academy of Sciences, which are scheduled for launch around the 2030s~\cite{LISA,Taiji}.
Each mission comprises a triangular configuration of three spacecraft (S/C) orbiting the Sun, linked by laser interferometers, and sensitive to the low-frequency band around millihertz.

Compared to the instantaneous GWs emitted by SBBHs in the merger phase, the low-frequency GWs emitted in the inspiral phase could last from several months to several hundred years, with many of them persisting longer than the mission duration  (assuming 4 years)~\cite{LISA_SOBBH}.
In contrast to the single chirp GW signal detected by ground-based GW detectors, GWs emitted by SBBHs during the mission duration are detected simultaneously and slowly increase in the frequency band of space-based GW detectors.
A minority of SBBHs are in close proximity to chirping, where the frequency rapidly increases before exiting the frequency band of the space-based GW detector~\cite{LISA_band}, which will subsequently enter the frequency band of ground-based GW detectors, enabling the observation of GW signals in multiband~\cite{multiband1,multiband2,multiband3}.
GW observations in the frequency band of space-based GW detectors can provide important information, such as the coalesce time and sky position of SBBHs, which is beneficial for ground-based GW detectors observations.
Performing multiband GW observations is of paramount importance as it contributes to the comprehension of the formation mechanism of SBBHs, the understanding of cosmic evolution, and the precise verification of modified gravity theories~\cite{test_GR,network}.
Hence, the observation of SBBHs using space-based GW detectors holds particular significance.
However, due to the lack of available data from space-based GW detectors at present, all relevant research is conducted based on simulations.
Therefore, understanding the population distribution and signal simulations of SBBHs is crucial for advancing our knowledge in this field.

In terms of the population model of SBBHs, the ancestors of SBBHs observed by LVK are potential sources of space-based GW detectors, and by analyzing the GW events, it becomes feasible to infer the population model of SBBHs~\cite{LIGO_population1,LIGO_population2,LIGO_population3}.
In Ref.~\cite{LISA_population}, Moore~\textit{et al}. assessed the impact of signal-to-noise ratio (SNR) thresholds on the astrophysical population of SBBHs observable by LISA, considering various merger rates and observation durations.
In Ref.~\cite{TianQin_SBBH}, Liu~\textit{et al}. utilized five different population models to estimate the expected number of detectable SBBHs and the accuracy of parameter estimation for TianQin by Fisher information matrix (FIM).
Furthermore, in Ref.~\cite{LISA_Taiji}, Chen~\textit{et al}. reported on the potential of utilizing both LISA and Taiji, along with their joint detection, for the detection of SBBHs.
Their work demonstrated the feasibility of future multiband GW observations, highlighting the possibilities and advantages of combining observations from multiple detectors.
Due to the variations arising from different models and theoretical calculations, significant differences may exist in the results presented across different references.
Meanwhile, our objective is not to delve into the discussion of those discrepancies.
Therefore, for the purpose of our study, we chose to utilize the four population models provided in Ref.~\cite{LIGO_population2} to establish our dataset, serving as the foundation for subsequent signal simulations.

In terms of the signal simulation of SBBHs, inaccurate waveforms can introduce deviations from the actual signals, leading to reduced detection rates, increased systematic errors, and even potential misinterpretation as effects beyond General Relativity (GR)~\cite{error1,error2,error3,error4}.
Additionally, incomplete removal of bright sources can impact the detection of other GW sources~\cite{LISA_waveform_model}.
Therefore, ensuring precise GW waveform simulations is crucial for reliable detection and accurate characterization of SBBHs and other sources in the GW data.
In the early inspiral phase, where the merger is still far away, GW waveform construction often relies on post-Newtonian (PN) formalism~\cite{PN_all}.
For instance, popular waveform models such as \texttt{EOBNR} and \texttt{IMRPhenomD} commonly utilize PN approximations to accurately simulate the waveform in inspiral phase~\cite{EOBNR1,EOBNR2,IMRPhenomD1,IMRPhenomD2}.
In Ref.~\cite{PN_accuracy}, Mangiagli~\textit{et al}. provided a preliminary estimation of the required PN model accuracy for SBBHs with LISA.
They derived the equations of motion (EOM) using numerical methods and utilized a Fourier-domain PN waveform from Ref.~\cite{fast_PN} to estimate the minimum waveform accuracy requirement.
Their results indicated that for 90\% of LISA-relevant SBBHs, a 2PN waveform accuracy is deemed adequate, while SBBHs with short time (i.e., $\tau\le4$ yr) require higher-order waveform accuracy.

We employ various methods to estimate accuracy requirement, with a particular focus on applications.
Compared to previous studies, our research not only considers different population models provided by LVK, but also investigates the relevant results in LISA, Taiji, and their joint detection.
By comparing factors such as mismatch, detection rate, and accuracy requirement, we provide a comprehensive assessment of the minimum accuracy required for SBBHs with space-based GW detectors.

This paper is organized as follows.
In Sec.~\ref{GW Signal}, we introduce the different PN models used in our paper, the four population models provided by LVK, and the necessary parameters for constructing the simulated dataset.
In Sec.~\ref{Detectors and TDI}, we review the relevant aspects of space-based GW detectors, including response functions, noise, and Time-Delay Interferometry (TDI).
In Sec.~\ref{Methodology}, we present the different methods employed for accuracy estimation, along with the basic parameter estimation.
In Sec.~\ref{Results}, we present the results of our simulations and analyses, providing accuracy calculations for different PN models.
Finally, we summarize our results in Sec.~\ref{Conclusion}.

\section{GW Signal}\label{GW Signal}
\subsection{Post-Newtonian models}\label{PN_models}
Most inspiraling SBBHs can be considered to possess nearly circular orbits, as the emission of GWs causes their orbital energies to decrease gradually, resulting in a chirp signal with progressively increasing frequency and amplitude~\cite{PN_all}.
SBBHs within the frequency band of space-based GW detectors are predominantly in the early inspiral phase, exhibiting minimal frequency variations.
Consequently, the GW waveforms can be expressed as follows~\cite{waveform}:
\begin{equation}\label{Eq:waveform}
\begin{aligned}
    h_{+}(t) & = \mathcal{A} \frac{1+\cos^{2}(\iota)}{2}\cos[\Phi(t)], \\
    h_{\times}(t) & = \mathcal{A} \cos(\iota)\sin[\Phi(t)], 
\end{aligned}
\end{equation}
with
\begin{equation}\label{Eq:wave_amp}
    \mathcal{A} =\frac{4}{D_L}\left(\frac{G\mathcal{M}_{\mathrm{c}}}{c^{2}}\right)^{5/3}\left(\frac{\pi f(t)}{c}\right)^{2/3},
\end{equation}
where $\mathcal{M}_{\mathrm{c}}=(m_1m_2)^{3/5}/(m_1+m_2)^{1/5}$ is the chirp mass, $m_1$ and $m_2$ are the masses
of SBBH, $D_L$ is the luminosity distance, $\iota$ is the inclination angle, $\Phi(t)=\int{2\pi f(t)}\mathrm{d} t$ is the GW phase, $f(t)$ is the GW frequency, $G$ and $c$ are the gravitational constant and the speed of light.

In the cosmological distance, accounting for cosmic expansion is essential when considering the propagation of GWs from the source to the detector.
The frequency and mass of the GW sources are influenced by the underlying cosmic expansion.
It can be demonstrated that by utilizing the cosmological redshift parameter $z$, the parameters associated with the source can be corrected by a factor of $(1+z)$.
The parameters related to the GW source are multiplied (or divided) by that factor, yielding the observed parameters in the GW detectors~\cite{waveform}.
In our paper, all parameters are based on the observed parameters.

In the PN formalism, the small remainder terms resulting from the PN expansion of EOM are typically denoted as $\mathcal{O} (1/c^n)$ and referred to as $n/2$ PN terms~\cite{PN_all}.
Through utilizing various PN approximations and considering possible effects, we can compute the energy flux $\mathcal{F} $ and total energy $E$. 
In case of a circular orbit, there is no need to employ the angular momentum balance equation or perform an averaging procedure~\cite{PN_SO}.
Instead, the evolution of frequency and phase can be obtained by solving the energy balance equation $\mathrm{d}E/\mathrm{d}t= -\mathcal{F}$.

The GW phase $\Phi$ and the GW frequency $f$ are related to the orbital phase $\phi$ and the angular frequency $\omega$ of SBBHs, which can be described as follows:
\begin{equation}
\Phi=2\phi,\quad \phi=\int{\omega}\mathrm{d}t,\quad f=\omega/\pi.
\end{equation}
For convenience, we employ dimensionless variables:
\begin{equation}
    x=\left(\frac{GM\omega}{c^3}\right)^{2/3},\quad \Theta=\frac{\nu c^3}{5GM}\tau,
\end{equation}
where $M=m_1+m_2$ is the total mass, $\nu=m_1m_2/(m_1+m_2)^2$ is the symmetric mass ratio, $\tau=t_c-t$ is the time to coalescence, and $t_c$ is the coalescence time.
For instance, in the lowest approximation, i.e., 0PN model, the orbital phase $\phi$ with initial phase $\phi_0$ can be writen as~\cite{PN_SO}
\begin{equation}
    \phi=\phi_0-\frac{x^{-5/2}}{32\nu},\quad x=\frac14\Theta^{-1/4}.
\end{equation}

In this paper, we consider the effects up to 3.5PN and incorporate the influence of spin effects.
We assume that the spins of the SBBHs are aligned with the orbital angular momentum, or equivalently, we neglect any components where the spins are misaligned with the orbital angular momentum.
That assumption ensures that the orbital plane and the spin direction remain fixed, i.e., non-precessing.
The different PN models discussed in the paper are all in the circular orbits and non-precessing. 
The impact of eccentricity and precession on waveforms is separately explained in Sec.~\ref{Eccentricity and precession}.

Typically, spin effects are divided into linear spin-orbit (SO) effects and quadratic spin-spin (SS) effects.
Therefore, the overall orbital phase can be expressed as a combination of non-spin (NS), SO, and SS contributions:
\begin{equation}\label{Eq:phase}
    \phi=\phi_0-\frac{x^{-5/2}}{32\nu}\sum_{p}(\varphi _p^{\mathrm{NS}}+\varphi _p^{\mathrm{SO}}+\varphi _p^{\mathrm{SS}})x^p,
\end{equation}
where the specific expressions of $x$ can be found in Eq.~(316) of Ref.~\cite{PN_all}, $\varphi _p^{\mathrm{NS}}$ and $\varphi _p^{\mathrm{SO}}$ can be found in Eq.~(16) and Eq.~(19) of Ref.~\cite{PN_SO}, and $\varphi _p^{\mathrm{SS}}$ can be found in Eq.~(14) of Ref.~\cite{PN_SS}.
Through the analytical expressions in Refs.~\cite{PN_all,PN_SO,PN_SS}, we can obtain the phase of different PN orders in Eq.~(\ref{Eq:phase}), and by substituting Eqs.~(\ref{Eq:waveform}) and~(\ref{Eq:wave_amp}), we can obtain the GW waveform in the time domain.

\begin{table}[ht]
\centering
\renewcommand{\arraystretch}{1.5}
\caption{The PN terms of NS, SO and SS. Checked items indicate non-zero, while unchecked items indicate zero.}\label{Tab:PN_model}
\begin{tabular*}{\columnwidth}{@{\extracolsep{\fill}}cccc@{}}
\hline
 PN & NS& SO & SS \\
\hline
0 & \checkmark & - &- \\
1 & \checkmark & - &- \\
1.5 & \checkmark & \checkmark & -\\
2 & \checkmark & - &\checkmark \\
2.5 & \checkmark & \checkmark &- \\
3 & \checkmark & \checkmark &\checkmark \\
3.5 & \checkmark & \checkmark &\checkmark \\
\hline
\end{tabular*}
\end{table}

\begin{figure}[ht]
    \begin{minipage}{\columnwidth}
        \centering
        \includegraphics[width=0.95\textwidth,
        trim=0 0 0 0,clip]{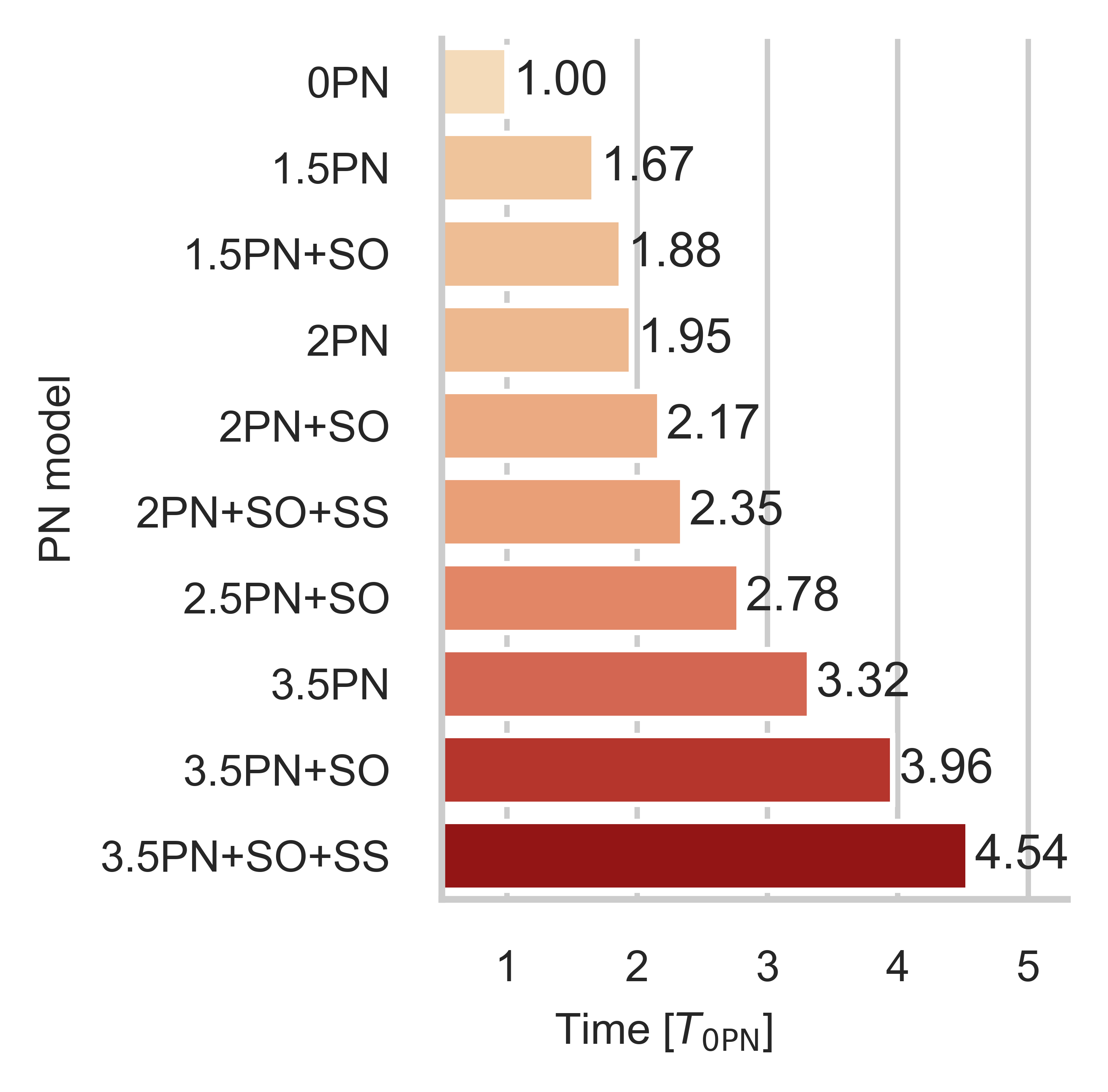}
        \caption{Comparison of time for generating GW signals using different PN models. The coordinates are expressed in units of $T_{\mathrm{0PN}}$, which is the time required to generate GW signals using 0PN model. The time sets represent the cumulative time taken to generate GW signals for hundreds of different parameters and durations. For each time set of GW signals, all parameters except the PN model remain the same.}\label{Fig:time}
    \end{minipage}
\end{figure}

\begin{figure*}[ht]
    \begin{minipage}{\textwidth}
        \centering
        \includegraphics[width=0.9\textwidth,
        trim=0 0 0 0,clip]{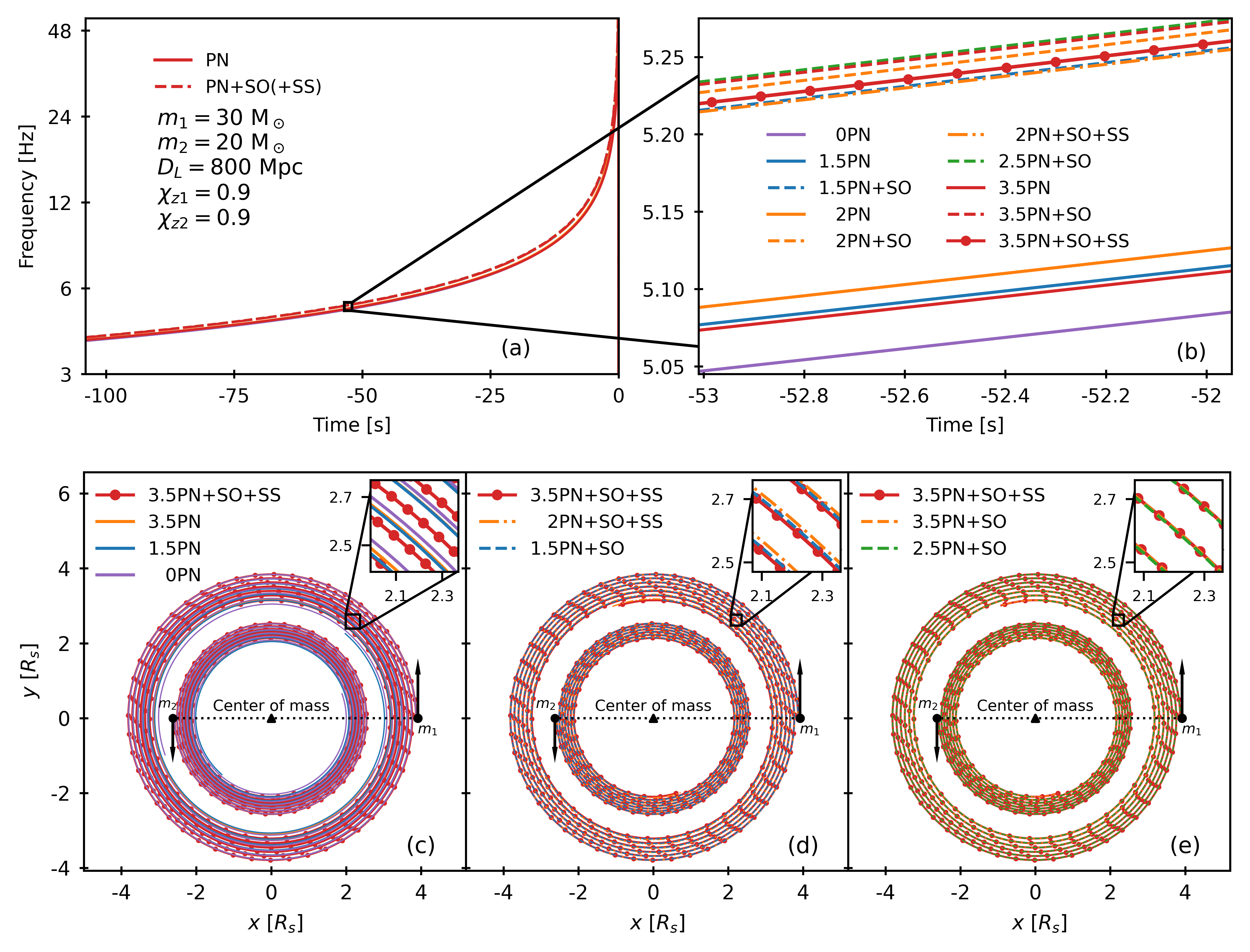}
        \caption{Comparison of frequency and orbit for different PN Models. \textit{Top panel}: (a) shows the variation of frequency with time for nearly 100 seconds before merger, with the parameters of the GW source listed in the figure. (b) is a magnified view of (a). those signal simulations are all based on the coalescence time $t_c=0$. \textit{Bottom panel}: The orbital evolution of the SBBHs, with the center of mass as the origin, exhibits variations in GW within the frequency range of 25 to 35~Hz. The coordinates are in units of Schwarzschild radii $R_s=2GM/c^2$. The SBBHs have the same initial positions, using the initial position based on 3.5PN+SO+SS model as a reference to compare the differences after a certain period of evolution.}\label{Fig:frequncy_orbit}
    \end{minipage}
\end{figure*}

The order of PN model, $p$, can be either an integer or a half-integer, and the non-zero PN terms are listed in Table~\ref{Tab:PN_model}.
For convenience, we denote 3.5NS+3.5SO+3.5SS as 3.5PN+SO+SS.
Based on Table~\ref{Tab:PN_model}, we select various PN models incorporating different PN orders and effects.
The corresponding generation times are detailed in Fig.~\ref{Fig:time}, which shows that an increase in PN order and associated effects leads to a proportional rise in the time required for simulating GW signals.
That result prompts a careful consideration of how to strike a balance between the distinctions among different PN models and the computational time involved.

For better demonstrating the distinctions among those PN models, we calculate the frequency variations before the merger and simulate the orbital evolution of the binary system using the Kepler's laws in Fig.~\ref{Fig:frequncy_orbit}.
The GW frequencies depicted in Fig.~\ref{Fig:frequncy_orbit} are not in the frequency band of space-based GW detectors, due to more distinct frequency variations and marginal differences among PN models in the period close to the coalescence which is not in millihertz band. 
Thus, to emphasize those distinctions, we specifically chose the GWs from SBBHs close to the coalescence, nevertheless for the subsequent signal simulations and accuracy calculations, we utilize the GWs from SBBHs within the frequency band of space-based GW detectors.

From Fig.~\ref{Fig:frequncy_orbit}, it is significant for the influence of spin effects, especially SO effect.
Considering the impact of spin effects, even lower-order PN models are more accurate than higher-order PN models, e.g., 1.5PN+SO model is more accurate than 3.5PN, with 3.5PN+SO+SS model as the standard reference.
Equation~(\ref{Eq:phase}) and References~\cite{PN_SO,PN_SS} indicate that NS and SO effects positively contribute to the phase and frequency, while SS effect has a negative contribution, e.g., the frequency of 2.5PN+SO+SS model is lower than that of the 2.5PN+SO model.
Meanwhile in the orbital evolution shown in Fig.~\ref{Fig:frequncy_orbit}, 2.5PN+SO model is more compatible with 3.5PN+SO+SS model than 2.5PN+SO+SS.
That suggests that in some cases, considering more effects may not lead to better results.
In Sec.~\ref{Accuracy of different PN models}, we provide a rigorous comparison of the accuracy among different PN models.

\subsection{Population models}\label{Population models}
\begin{figure*}[ht]
    \begin{minipage}{\textwidth}
        \centering
        \includegraphics[width=0.9\textwidth,
        trim=0 0 0 0,clip]{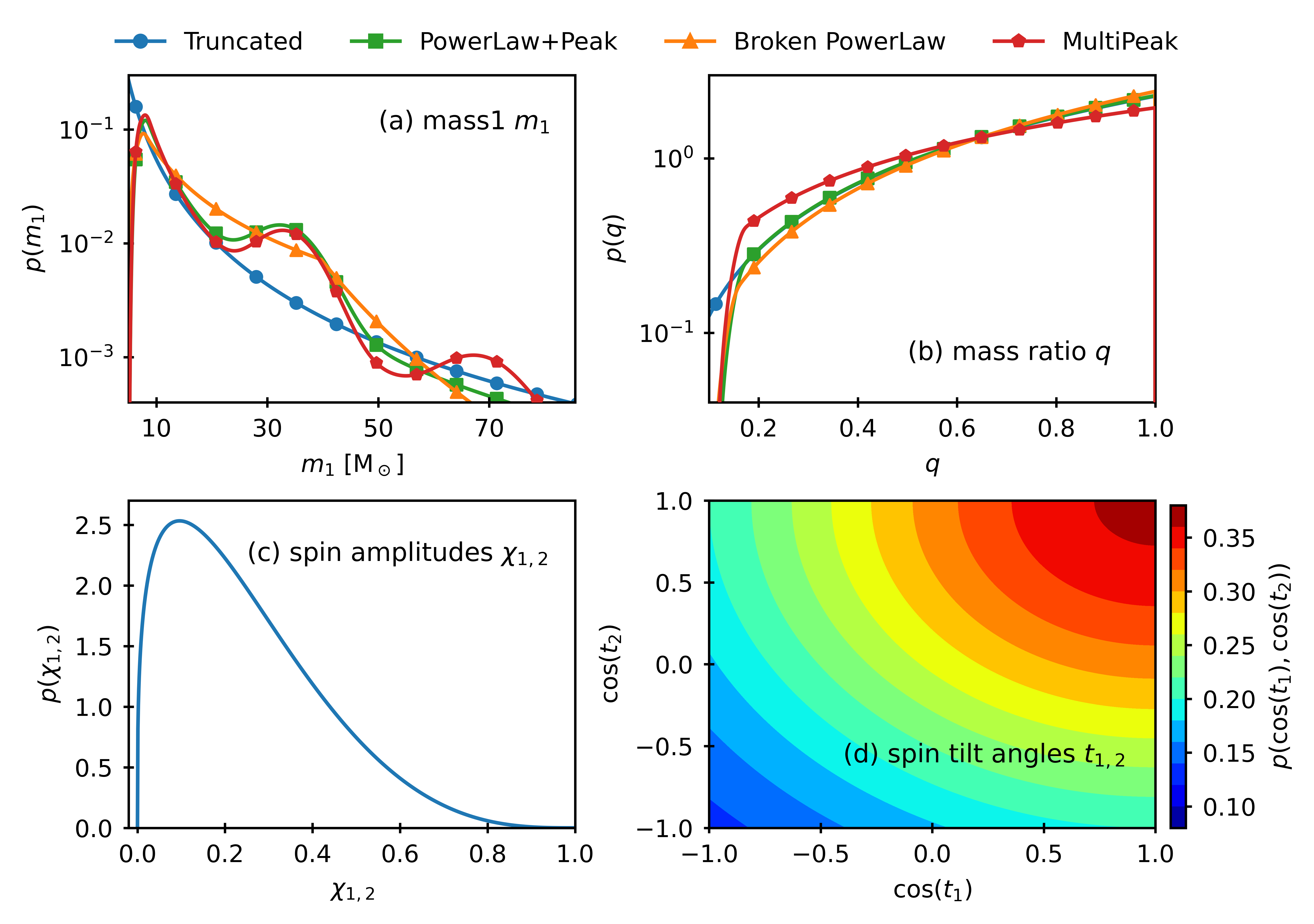}
        \caption{Distribution of mass and spin parameters. \textit{Top panel}: (a) shows comparison of four types of SBBH mass distribution models: \texttt{Truncated}, \texttt{PowerLaw+Peak}, \texttt{Broken PowerLaw}, and \texttt{MultiPeak}~\cite{LIGO_population1}. (b) is the mass ratio distribution of those four population models. \textit{Bottom panel}: (c) and (d) are the amplitude and tilt angle distributions of \texttt{DEFAULT} spin model~\cite{LISA_SOBBH}.}\label{Fig:population}
    \end{minipage}
\end{figure*}
To improve the realism of our simulations, we employ various population models for SBBHs to generate diverse datasets.
We focus on four critical aspects: the mass distribution, the spin distribution, the redshift evolution, and the distribution of other parameters.
Those aspects collectively contribute to a more comprehensive and accurate description of the observed SBBH population.

In our study, we explore four different models for the mass distribution, and each offers a different method to describe the mass distribution of SBBHs.
Here is a brief description of those models~\cite{LIGO_population1,LIGO_population2,LIGO_population3}:
\begin{enumerate}
    \item \texttt{Truncated}: The simplest mass model considered is characterized by a power-law distribution for the primary masses, with hard cutoffs at both low and high masses. This model utilizes only four parameters to describe the mass distribution effectively.
    \item \texttt{PowerLaw+Peak}: Compared to the previous model, this one incorporates additional features to account for the low-mass and high-mass ends of the distribution. It includes a smoothing function at low masses and introduces a Gaussian peak at high masses. Therefore, this model consists of eight parameters to fully describe the mass distribution.
    \item \texttt{Broken PowerLaw}: This model consists of two power-law distributions with different exponents, connected by a break in the middle. It incorporates a smoothing function at low masses and is described by seven parameters to characterize the mass distribution.
    \item \texttt{MultiPeak}: Similar to \texttt{PowerLaw+Peak}, this model includes two Gaussian peaks. Thus, the model has eleven parameters.
\end{enumerate}

The specific mathematical definitions and parameter values for the four mass distribution models mentioned can be found in Appendix~B of Ref.~\cite{LIGO_population1}.
In Fig.~\ref{Fig:population} (a) and (b), we present the normalized mass distribution of those four models. Our subsequent simulations are based on those distributions.

For the spin distribution of SBBHs, we use the simplest \texttt{DEFAULT} spin model~\cite{LISA_SOBBH}.
In this model, the dimensionless spin amplitude distribution follows a Beta distribution, while the tilt angle of the spin is described by a combination of the Isotropic distribution and the Gaussian distribution.
Fig.~\ref{Fig:population} (c) and (d) shows the distribution of spin amplitude and tilt angle.

For the redshift evolution of the SBBH merger rate, we adopt the \texttt{NONEVOLVING} model, assuming a constant merger rate density $\mathcal{R}(z)$ that does not vary with redshift $z$~\cite{LIGO_population1}.
For \texttt{PowerLaw+Peak}, \texttt{Broken PowerLaw} and \texttt{MultiPeak}, after incorporating the constraints from the LVK observations, the derived $\mathcal{R}$ values are very close, and we set $\mathcal{R}=23.9\ \mathrm{Gpc}^{-3}\mathrm{yr}^{-1} $.
Additionally, \texttt{Truncated} has a slightly higher $\mathcal{R}$ value than the other models, but its results remain consistent within the statistical uncertainties.
Hence, we set $\mathcal{R}=33\ \mathrm{Gpc}^{-3}\mathrm{yr}^{-1} $ for \texttt{Truncated}.

\begin{table}[ht]
\centering
\renewcommand{\arraystretch}{1.5}
\caption{Parameter distribution used in population production~\cite{LISA_SOBBH}. $U[a,b]$ represents a uniform distribution from $a$ to $b$.}\label{Tab:parameters}
\begin{tabular*}{\columnwidth}{@{\extracolsep{\fill}}cc@{}}
\hline
 Parameter & Distribution \\
\hline
Time To Coalescence $\tau$ & $U[0,\tau_\mathrm{max}/(1+z)]$ yrs\\
Ecliptic Longitude $\lambda$ & $U[0,2\pi]$ rad\\
Ecliptic Latitude $\beta$ & $\arcsin(U[-1,1])$ rad\\
Inclination $\iota$ & $\arccos(U[-1,1])$ rad\\
Polarization $\psi$ & $U[0,2\pi]$ rad\\
Initial Phase $\phi_0$ & $U[0,2\pi]$ rad\\
\hline
\end{tabular*}
\end{table}

The additional parameters for SBBHs, including the time to coalescence $\tau$, sky positions $\lambda$ and $\beta$, inclination $\iota$, polarization $\psi$, and initial phase $\phi_0$, are generated based on the prior data from Table~\ref{Tab:parameters}.
With those known parameters, we can calculate the remaining parameters, such as the initial frequency $f_0$ and luminosity distance $D_L$, which can be written as~\cite{waveform}
\begin{equation}\label{Eq:f0}
    f_0=\frac1{8\pi}\left[\frac15\left(\frac{G\mathcal{M}_c}{c^3}\right)^{5/3}\tau\right]^{-3/8},
\end{equation}
\begin{equation}\label{Eq:DL}
    D_L=\frac{c(1+z)}{H_0} \int_0^z{\frac{\mathrm{d}z'}{\sqrt{\Omega_m(1+z')^3+\Omega_\Lambda } } },
\end{equation}
where we adopt the $\mathrm{\Lambda CDM}$ cosmological model with parameters derived from the "\textit{Planck} 2018 results"~\cite{Planck_2018}, corresponding to the Hubble constant $H_0=67.37\ \mathrm{km}\ \mathrm{s}^{-1} \mathrm{Mpc}^{-1}$, matter density parameter $\Omega_m =0.315 $, and dark energy density parameter $\Omega_\Lambda  =0.685$.

To estimate the potential number of SBBHs detectable by space-based GW detector, we utilize a calculation method similar to the one described in Refs.~\cite{multiband1,LISA_band}.
The calculation for the number of SBBHs can be expressed as
\begin{equation}\label{Eq:data_number}
    N=\int{\mathcal{R}(z)p(\boldsymbol{\theta} )\frac{\mathrm{d}V_c(z)}{\mathrm{d}z}\frac{1}{1+z}} \mathrm{d}z\mathrm{d}\boldsymbol{\theta} \mathrm{d}\tau,
\end{equation}
where $\boldsymbol{\theta}$ collectively denotes different parameters, $p(\boldsymbol{\theta} )$ is their probability density function, $\mathrm{d}V_c(z)$ is the comoving volume, and $1/(1+z)$ accounts for the cosmic expansion.
In our study, we limit the distance of the GW sources to the range $10^{-3}\le z\le 2$, and set the upper limit for $\tau_\mathrm{max} $ to be 100 yrs.

Using Eq.~(\ref{Eq:data_number}), we can calculate the variation and total number of the SBBH with redshift.
We use \texttt{NumPy}~\cite{numpy} for weight sampling to generate samples that conform to the parameter distribution in Fig.~\ref{Fig:population} and Table~\ref{Tab:parameters}.
Using Eqs.~(\ref{Eq:f0}) and~(\ref{Eq:DL}) to calculate the remaining parameters, we can obtain parameter information containing several SBBH sources and construct a dataset that conforms to the population model.
Based on the above, we construct four different datasets representing different SBBH population models, incorporating various parameters for the GW sources.
In Sec.~\ref{Detection number}, we present a detailed analysis and comparison of the results for high SNR sources in those datasets.

\section{Detectors and TDI}\label{Detectors and TDI}
\subsection{Detector’s response}

\begin{figure}[ht]
    \begin{minipage}{\columnwidth}
        \centering
        \includegraphics[width=0.9\textwidth,
        trim=0 0 0 0,clip]{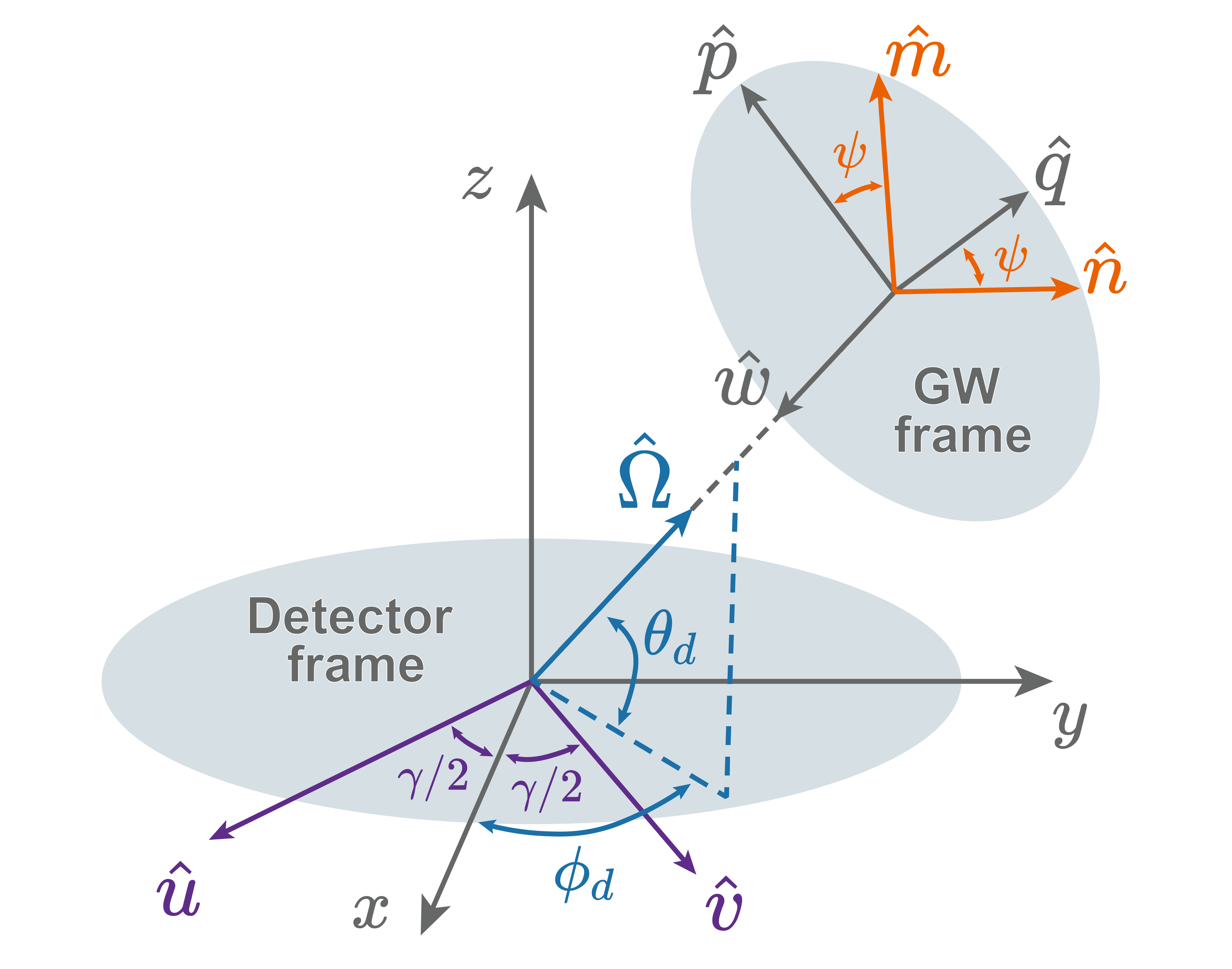}
        \caption{Relationship between detector frame and GW frame.}\label{Fig:frame}
    \end{minipage}
\end{figure}

LISA and Taiji both consist of a triangular configuration of three S/C and can be regarded as three dual-arm Michelson laser interferometers.
They detect GWs by measuring the relative changes in the lengths of the two arms, and to describe the response of the detectors to a GW, we employ the method in Ref.~\cite{response}.
As shown in Fig.~\ref{Fig:frame}, the detector frame is established using a set of orthogonal unit vectors \{$\hat{x}, \hat{y}, \hat{z}$\}, while the GW frame is constructed using \{$\hat{p}, \hat{q}, \hat{w}$\}, with $\hat{w}$ representing the propagation direction of the GW, and the unit vector $\hat{\Omega}=-\hat{w}$ denoting the position of the GW source.
For the detector, the two arms have an angle $\gamma =60^\circ$, and the unit vectors $\hat{u}$ and $\hat{v}$ representing the arms are positioned on opposite sides of the $x$-axis.
For the GW, there exists an additional rotational degree of freedom that can be fixed by specifying the polarization $\psi$.
That allows us to determine the GW polarization states using \{$\hat{m}, \hat{n}, \hat{w}$\}.

The detector frame and GW frame can be interchanged through a Euler rotation, and all of those concepts can be expressed using the observation angles of the detector, $\phi_d$ and $\theta_d$.
The relationship between the GW signal in the detectors and the polarization states of the GW can be expressed as
\begin{equation}\label{Eq:GW+singal}
    h(t)=F^\times h_\times(t)+F^+h_+(t),
\end{equation}
where $h_\times(t)$ and $h_+(t)$ are given in Eq.~(\ref{Eq:waveform}), the antenna response functions $F^\times$ and $F^+$ are defined as
\begin{equation}
    F^\times=D^{ij}e_{ij}^\times,\quad F^+=D^{ij}e_{ij}^+.
\end{equation}
We can define the polarization tensors using \{$\hat{m}, \hat{n}, \hat{w}$\} as follows:
\begin{equation}
    e_{ij}^+=\hat{m}_i\hat{m}_j-\hat{n}_i\hat{n}_j,\quad e_{ij}^\times=\hat{m}_i\hat{n}_j+\hat{n}_i\hat{m}_j.
\end{equation}
And the detector tensor $D^{ij}$ can be represented using transfer function $\mathcal{T}$~\cite{transfer_function}:
\begin{equation}
    D^{ij}=\frac{1}{2}[\hat{u}^i\hat{u}^j\mathcal{T}(f,\hat{u}\cdot\hat{w})-\hat{v}^i\hat{v}^j\mathcal{T}(f,\hat{v}\cdot\hat{w})],
\end{equation}
with 
\begin{widetext}
\begin{equation}\label{Eq:transfer_function}
    \mathcal{T}(f,\hat{a}\cdot\hat{w})= \frac{1}{2}\left\{\operatorname{sinc}\left[\frac{f}{2f_{*}}(1-\hat{a}\cdot\hat{w})\right]\exp\left[-i\frac{f}{2f_{*}}(3+\hat{a}\cdot\hat{w})\right]+\mathrm{sinc}\left[\frac{f}{2f_{*}}(1+\hat{a}\cdot\hat{w})\right]\exp\left[-i\frac{f}{2f_{*}}(1+\hat{a}\cdot\hat{w})\right]\right\}, 
\end{equation}
\end{widetext}
where $\operatorname{sinc}(x)=\sin x/x$, $f_{*}=c/(2\pi L)$ is the transfer frequency and $L$ is the arm length of the detector.
The arm length for LISA $L=2.5\times10^9$ m, while for Taiji, the arm length $L=3\times10^9$ m.
The transformations between the detector coordinate system ($\phi_d,\theta_d$) and the ecliptic coordinate system ($\lambda,\beta$) can be found in Refs.~\cite{response,my_paper}.

Additionally, for a space-based GW detector around the Sun, the periodic motion will produce the Doppler phase, which is given by~\cite{Doppler_phase,my_paper}
\begin{equation}\label{Eq:Doppler}
    \Phi_{D}(t) = 2\pi f(R/c)\cos\beta\cos(2\pi f_mt-\lambda ),
\end{equation}
where $R$ = 1 A.U. is the distance between the Sun and the Earth, $f_m$ = 1/year is the geocentric orbit modulation frequency and ($\lambda,\beta$) are the ecliptic coordinates of the GW source.
During the simulation of GW signals, we consider the influence of the Doppler effect by incorporating the Doppler phase $\Phi_D(t)$ from Eq.~(\ref{Eq:Doppler}) into the phase $\Phi(t)$ in Eq.~(\ref{Eq:waveform}). 
That ensures the Doppler effect is properly accounted for in the simulated signals.

\subsection{Time-Delay Interferometry and noise}\label{TDI_noise}

For space-based GW detectors, TDI is essential to suppress laser frequency noise and achieve detection goals~\cite{TDI1,TDI2,TDI3}.
Different combinations of TDI channels yield varying responses to GW signals and instrument sensitivities~\cite{different_TDI}.

\begin{figure}[ht]
    \begin{minipage}{\columnwidth}
        \centering
        \includegraphics[width=0.9\textwidth,
        trim=0 0 0 0,clip]{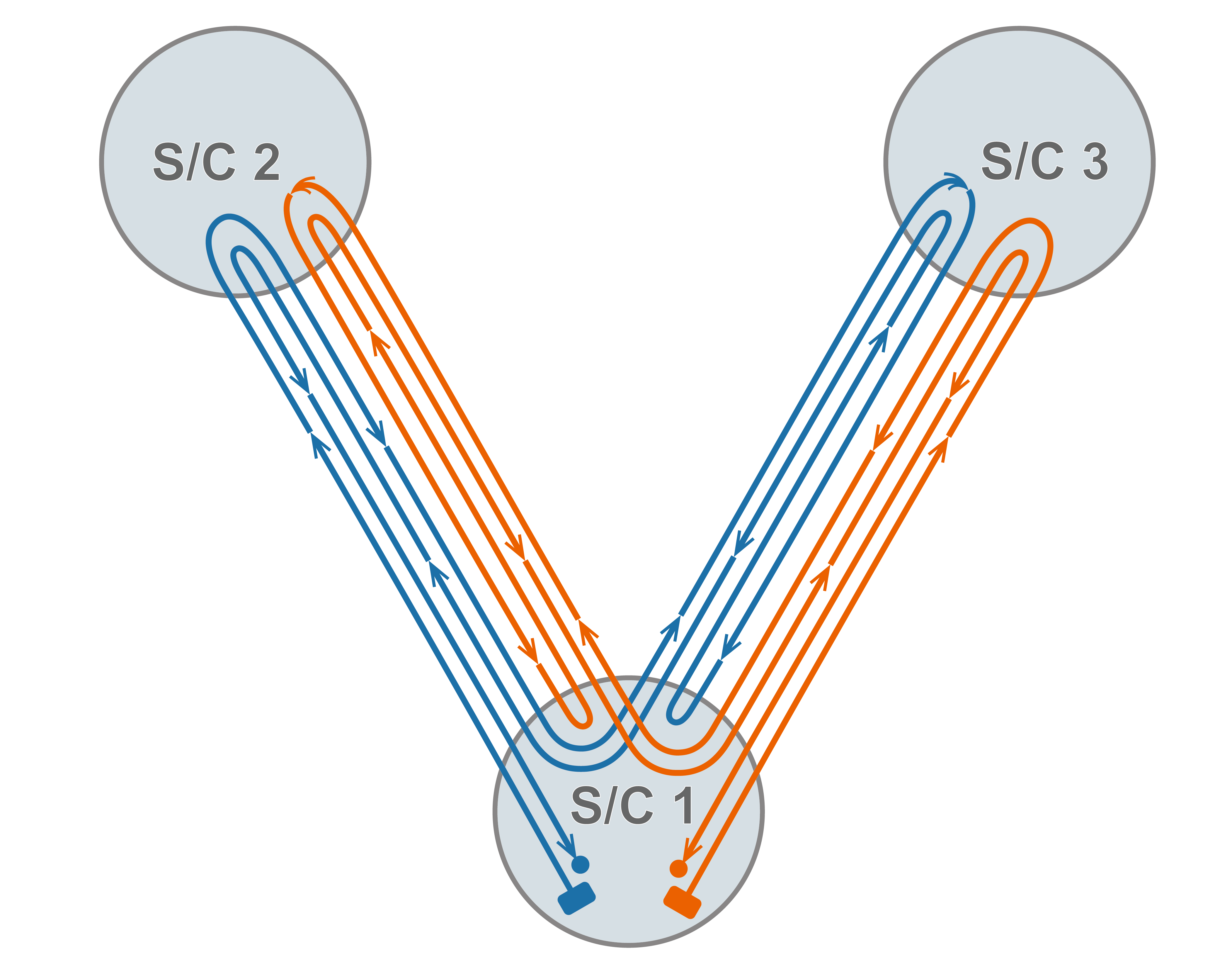}
        \caption{Michelson X channel of the 2nd generation TDI.}\label{Fig:TDI}
    \end{minipage}
\end{figure}
We use the second generation Michelson XYZ channels, and the laser path for the X channel is shown in Fig.~\ref{Fig:TDI}.
Based on the laser path, the expression for the X-TDI 2.0 can be described as~\cite{LISA_noise1}
\begin{equation}\label{Eq:X-TDI}
\begin{aligned}
    X_{2.0}  = &\eta_{13}+D_{13}\eta_{31}+D_{13}D_{31}\eta_{12}+D_{13}D_{31}D_{12}\eta_{21} \\
    +&D_{13}D_{31}D_{12}D_{21}\eta_{12}+D_{13}D_{31}D_{12}D_{21}D_{12}\eta_{21} \\
    +&D_{13}D_{31}D_{12}D_{21}D_{12}D_{21}\eta_{13}\\
    +&D_{13}D_{31}D_{12}D_{21}D_{12}D_{21}D_{13}\eta_{31}\\
    -&(\eta_{12}+D_{12}\eta_{21}+D_{12}D_{21}\eta_{13}+D_{12}D_{21}D_{13}\eta_{31}\\
    +&D_{12}D_{21}D_{13}D_{31}\eta_{13}+D_{12}D_{21}D_{13}D_{31}D_{13}\eta_{31} \\
    +&D_{12}D_{21}D_{13}D_{31}D_{13}D_{31}\eta_{12}\\
    +&D_{12}D_{21}D_{13}D_{31}D_{13}D_{31}D_{12}\eta_{21} ) ,
\end{aligned}
\end{equation}
where $D_{ij}$ is the time-delay operator, $D_{ij}\eta(t)=\eta(t-L_{ij}/c)$, $\eta_{ij}$ are the combined observables from $\mathrm{S/C}j$ to $\mathrm{S/C}i$, $L_{ij}$ is the arm length from $\mathrm{S/C}i$ to $\mathrm{S/C}j$~\cite{Alternative_LISA-TAIJI,TDI4}.
The expressions for Y-TDI 2.0 and Z-TDI 2.0 are in the same form as Eq.~(\ref{Eq:X-TDI}), with the notation of $\mathrm{S/C}i$ and $\mathrm{S/C}j$ being interchanged.

Through employing TDI, we assume that laser frequency noise is adequately suppressed and significantly lower than other kinds of noise.
Consequently, we only focus on the instrumental noise of space-based GW detectors, which can be primarily composed of two components: a low-frequency noise component, characterized by acceleration noise, and a high-frequency noise component, characterized by displacement noise~\cite{LISA_noise1,LISA_noise2,Taiji_noise1,Taiji_noise2}.
LISA and Taiji have the same acceleration noise, described as
\begin{equation}
    \sqrt{S^{a}}=\frac{3\times10^{-15}}{2\pi fc}\sqrt{1+\left(\frac{0.4\ \mathrm{mHz} }{f}\right)^{2}}\sqrt{1+\left(\frac{f}{8\ \mathrm{mHz}}\right)^{4}}. 
\end{equation}
The displacement noise for LISA can be expressed as
\begin{equation}\label{Eq:dis_noise_LISA}
    \sqrt{S^{d}}=15\times10^{-12}\times\frac{2\pi f}{c}\sqrt{1+\left(\frac{2\ \mathrm{mHz}}{f}\right)^4}, 
\end{equation}
while for Taiji, the displacement noise is
\begin{equation}\label{Eq:dis_noise_Taiji}
    \sqrt{S^{d}}=8\times10^{-12}\times\frac{2\pi f}{c}\sqrt{1+\left(\frac{2\ \mathrm{mHz}}{f}\right)^4}.
\end{equation}
Considering the effects of acceleration noise $\sqrt{S^{a}}$ and displacement noise $\sqrt{S^{d}}$, we can derive the power spectral density (PSD) of the noise and the sensitivity curve of the detector~\cite{LISA_noise1}.
In Bayesian statistical inference, we use the PSD and GW response after TDI 2.0, while in other parts, we utilize the non-sky-averaged PSD based on the sensitivity curve mentioned in Ref.~\cite{non_sky_averaged_PSD}.
That is done to maintain consistency with the Bayesian parameter estimation in Ref.~\cite{PE_LISA} and the SBBHs detection in Ref.~\cite{LISA_Taiji}.
Given that both definitions are valid and serve as suitable calculation conditions, and considering the focus of our paper is to compare and apply different PN models, it is appropriate to exclude the discussion of the differences between the two definitions in our study.

\section{Methodology}\label{Methodology}
\subsection{Mismatch and SNR}
To compare the differences between different PN models, we consider the 3.5PN+SO+SS model waveform as the exact physical waveform, denoted as $h_e$, and the waveforms from different PN models as $h_m$.
A useful measure of how close the two waveforms are to each other is given by Faithfulness $F$ (also known as Match), that is, the overlap maximized only over the coalescence time $t_c$ and phase $\phi_c$ of the waveform~\cite{overlap},
\begin{equation}\label{Eq:overlap_fun}
    F(h_e,h_m) =\max_{\phi_{c},t_{c}}\frac{\left \langle h_e| h_m \right \rangle }{\sqrt{\left \langle h_e| h_e \right \rangle \left \langle h_m| h_m \right \rangle} } ,
\end{equation}
with the inner product $\left \langle \cdot |\cdot \right \rangle$ defined as
\begin{equation}
    \left \langle a |b \right \rangle=4\text{Re}\int_0^{+\infty}{\frac{\tilde{a}^*(f)\tilde{b}(f)}{S_n(f)} }\mathrm{d} f,
\end{equation}
where $\tilde{a}(f)$ and $\tilde{b}(f)$ are the Fourier transformations of $a(t)$ and $b(t)$, $S_n(f)$ is the one-sided noise PSD.
The value of $F$ reflects the degree of similarity between $h_e$ and $h_m$.
For two waveforms that are relatively close, we often prefer to use the complementary measure, known as the Mismatch $\mathcal{MM}$, which can be expressed as
\begin{equation}\label{Eq:mismatch}
    \mathcal{MM}(h_e,h_m)=1-F(h_e,h_m).
\end{equation}

In the matched filtering, a bank of discrete modeled waveforms is used as filters~\cite{overlap}.
The optimal value for recovering SNR $\rho_{\mathrm{opt} }$ can be simply expressed as
\begin{equation}\label{Eq:SNR_opt}
    \rho _{\mathrm{opt} }=\sqrt{\left \langle h_e|h_e \right \rangle } ,
\end{equation}
and with a finite bank of filter templates, the recovered SNR $\rho$ is expressed as~\cite{PN_accuracy}
\begin{equation}\label{Eq:SNR_overlap}
    \rho=\frac{\left \langle h_e| h_m \right \rangle }{\sqrt{\left \langle h_m| h_m \right \rangle} }\approx F(h_e,h_m) \rho _{\mathrm{opt} }.
\end{equation}

To evaluate the effective loss of SNR caused by imperfect waveform models, measurement indicators such as $F$ or $\mathcal{MM}$ are commonly used.
Apparently, less accurate waveforms lead to lower $F$, resulting in a decrease of SNR for a specific GW signal and potentially impeding signal detection.
From Eq.~(\ref{Eq:SNR_overlap}), if the used model deviates from the actual GW signal by $F$, SNR will also experience a loss of $F$.
For high SNR sources, such as massive black hole binary, even with some SNR loss, the detection rate remains almost unaffected~\cite{MBHB}.
However, for relatively quiet sources like SBBH, the impact on the detection rate is significant, resulting in a decrease in the number of detections by a factor of $F^3$~\cite{LISA_waveform_model}.
Therefore, the accuracy of the waveform model is crucial for the detection rate of SBBHs, and improving the accuracy of the waveform model becomes essential for maximizing the chances of detecting SBBHs.

\subsection{Accuracy requirements}
In addition to considering the diversity between different PN models, accuracy requirement is equally important when using them in GW data analysis.
Setting appropriate accuracy requirement ensures that the scientific information can be extracted from GW data completely, while avoiding unnecessary computational burden with overly stringent accuracy during simulations.

In GW data processing, we approximate the parameter estimation of fixed SBBHs for individual sources, which means when we analyze a specific GW source, the influence from other GW sources is not considered.
In such case, there are several methods to assess accuracy requirement for different waveforms.
A simple and conservative approach for waveform accuracy requirement, which aims at avoiding the introduction of systematic biases comparable to the statistical uncertainties of individual sources, can be expressed by imposing constraints on the relationship between $\mathcal{MM}$ and the optimal SNR $\rho_{\mathrm{opt}}$~\cite{LISA_waveform_model}:
\begin{equation}\label{Eq:Accuracy1}
    \mathcal{MM}< 1/\rho^2 _{\mathrm{opt} }.
\end{equation}

In addition, we also adopt another simple yet strict requirement described in Ref.~\cite{Accuracy_requirements}.
The difference between two waveforms, $\delta h=h_m-h_e$, can be assessed using FIM.
If the standard deviation $\sigma$ calculated from FIM is greater than 1, the waveforms are considered indistinguishable.
That condition can be equivalently written as
\begin{equation}\label{Eq:Accuracy2}
    \left \langle \delta h|\delta h \right \rangle <1.
\end{equation}
The accuracy requirement described in Eq.~(\ref{Eq:Accuracy2}) establishes a strict requirement for measurements.
If a waveform satisfies that requirement, it implies that the two waveforms are indistinguishable to the detector. 
It represents an ideal requirement where more precise waveforms beyond that requirement would not result in further improvement in scientific measurements.
Conversely, less accurate waveforms may degrade certain measurements.
In Sec.~\ref{Accuracy of different PN models}, we discuss the accuracy requirement of different PN models, particularly in terms of detection rates and measurements, which primarily involve three equations: Eqs.~(\ref{Eq:SNR_overlap}), (\ref{Eq:Accuracy1}) and (\ref{Eq:Accuracy2}).

\subsection{Statistical inference}\label{Statistical inference}
In our paper, we aim to simplify the investigation of the bias generated by different PN models in statistical inference. 
Therefore, we offer a concise overview of the relevant concepts without delving into excessive detail.

Based on Refs.~\cite{PE_LISA,PE_TianQin,PE}, our statistical inference is performed using the Bayesian framework, where the posterior probability distribution $p(\boldsymbol{\theta}|d)$ with different parameters $\boldsymbol{\theta}$ is derived according to Bayes theorem:
\begin{equation}\label{Eq:Bayes_theorem}
    p(\boldsymbol{\theta}|d)\propto\mathcal{L}\left(d|\boldsymbol{\theta}\right)p(\boldsymbol{\theta}),
\end{equation}
where $p(\boldsymbol{\theta})$ represents the prior information.
For the Bayesian inference, we employ the likelihood function $\mathcal{L}(d|\boldsymbol{\theta})$ defined as~\cite{PE_LISA}
\begin{equation}\label{Eq:likelihood}
    \ln\mathcal{L}(d|\boldsymbol{\theta})=-\sum_k\frac{\langle d_k-h_k(\boldsymbol{\theta})|d_k-h_k(\boldsymbol{\theta})\rangle_k}2+\mathrm{const},\end{equation}
where $k$ stands for different TDI channels \{$k=\mathrm{X,Y,Z} $\}, $d$ is TDI output, $h(\boldsymbol{\theta})$ is PN waveform.
When the prior $p(\boldsymbol{\theta})$ is specified, we can employ Markov Chain Monte Carlo (MCMC) method to sample and infer the posterior $p(\boldsymbol{\theta}|d)$ from Eq.~(\ref{Eq:Bayes_theorem}).
In our paper, we utilize a specific affine-invariant ensemble sampler \texttt{emcee} for sampling and inference~\cite{emcee}.

\begin{figure*}[ht]
    \begin{minipage}{\textwidth}
        \centering
        \includegraphics[width=0.92\textwidth,
        trim=0 0 0 0,clip]{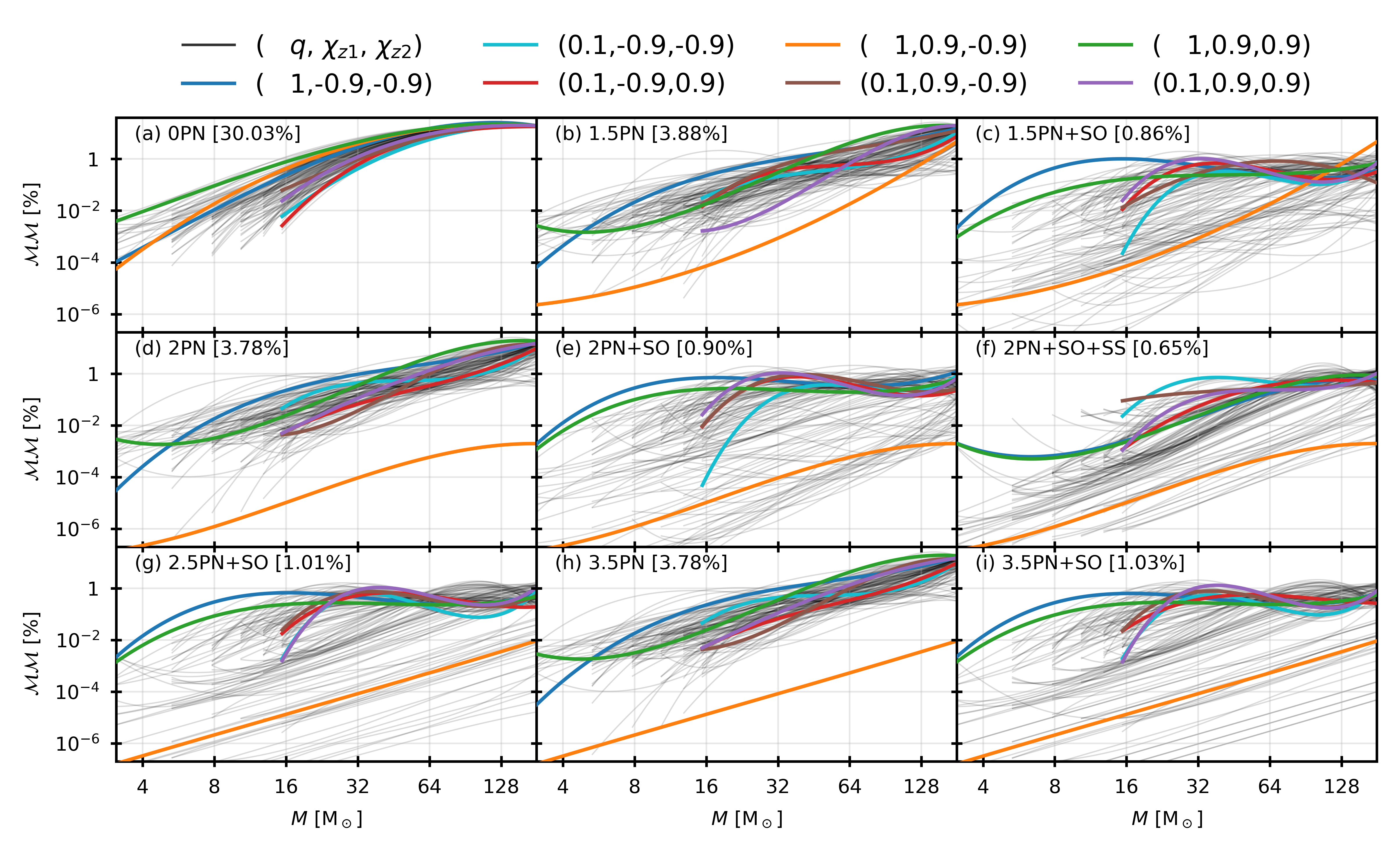}
        \caption{Comparison of mismatches $\mathcal{MM}$ for different PN models. The legend at the top represents different mass ratios $q=m_2/m_1<1$ and spins $\chi_{z1/2}$. The titles of each subplot, enclosed in brackets, indicate the average value of $\mathcal{MM}$. The curves in the figure are obtained by fitting a third-order polynomial in logarithmic space. Our calculation is based on the fitted curve, using 1000 points for each curve and calculating the proportion of points on all curves. The black curves represent the results under different parameters, while the colored curves represent the results under extreme parameters.}\label{Fig:mismatch}
    \end{minipage}
\end{figure*}

\section{Results}\label{Results}
\subsection{Mismatch}\label{Mismatch}
We simulate GW signal $h_m$ using different PN models, with the 3.5PN+SO+SS model $h_e$ as the reference.
The mismatches $\mathcal{MM}$ in LISA are calculated using Eqs.~(\ref{Eq:overlap_fun}) and (\ref{Eq:mismatch}).
The initial frequency of all SBBHs is set to be 8~mHz, a highly sensitive frequency for LISA.
The total mass $M$ ranged from $M_{\mathrm{min} } $ to $200\ \mathrm{M_\odot } $, assuming that the total mass of the SBBHs should not be less than $M_{\mathrm{min} }\ge (1+1/q)M_{\mathrm{NS} }$, where $M_{\mathrm{NS} }\sim 1.4\ \mathrm{M_\odot } $ is the mass of a neutron star~\cite{M_min}.
We systematically vary the mass ratio $q$ and spin $\chi_{z1/2}$, considering different scenarios. 
Specifically, we investigate the $\mathcal{MM}$ curves for different $q$ and $\chi_{z1/2}$: $q\in [0.1,1]$, $\chi_{z1/2}\in [-0.9,0.9]$.
The $\mathcal{MM}$ values for different PN models are computed based on the above parameter space, and the results are depicted in Fig.~\ref{Fig:mismatch}.

The results depicted in Fig.~\ref{Fig:mismatch} clearly demonstrate the significant advantage of incorporating spin effects in the PN models.
The average value of $\mathcal{MM}$ for the spin-inclusive models is only around 1\%.
Based on our calculations, for cases ranging from 0.26\% to 0.41\%, $\mathcal{MM}\ge0.6\%$, indicating that cases exceeding 99.6\%, the Match exceeds 0.994 (cf. Ref.~\cite{PN_accuracy}).
From an overall perspective, there is a clear increasing trend in $\mathcal{MM}$ values as the total mass $M$ increases.
In Fig.~\ref{Fig:population}, the SBBH mass distribution $p(m_1)$ tends to decrease with higher mass $m_1$.
Consequently, for the population, SBBHs with a larger proportion of lower masses exhibit relatively smaller $\mathcal{MM}$.

In Fig.~\ref{Fig:mismatch}, for PN models of the same order, such as 1.5PN and 1.5PN+SO models, there is an orange line that looks the same.
That is because that line represents a special case where a binary system has equal masses ($q=1$) and opposite spins ($\chi_{z1}=0.9,\ \chi_{z2}=-0.9$).
According to Eq.~(\ref{Eq:phase}) and Refs.~\cite{PN_SO,PN_SS}, when a binary system consists of equal masses with precisely opposite spins, the effects of SO/SS will be offset ($\varphi _p^{\mathrm{SO/SS}}=0$), resulting in no contribution to the phase $\phi$.
Consequently, PN models of the same order yield exactly the same results for such configurations.

An meaningful result from Fig.~\ref{Fig:mismatch} is that higher-order PN models do not necessarily yield smaller $\mathcal{MM}$ values than lower-order PN models.
For instance, the average $\mathcal{MM}$ for the 1.5PN+SO model is actually smaller than that for the 3.5PN+SO model.
That phenomenon also occurs in the frequency variation depicted in Fig.~\ref{Fig:frequncy_orbit}(b), which has been analyzed in the final paragraph of Sec.~\ref{PN_models}.
Hence, when simulating GW signals, blindly using high-order PN models will increase computational resource consumption and may not achieve the accuracy achieved by low-order PN models.
The above discussion is related to the overall analysis, and in Sec.~\ref{Accuracy of different PN models}, we investigate the accuracy of different PN models for detectable SBBHs.

\subsection{Detection number and SNR}\label{Detection number}
For the datasets, we employ the method described in Sec.~\ref{Population models}, which results in four different datasets.
Then we utilize Eq.~(\ref{Eq:SNR_opt}) to calculate the optimal SNR $\rho_{\mathrm{opt} }$ and identify SBBHs with $\rho_{\mathrm{opt} }$ greater than SNR threshold $\rho_{\mathrm{thr} }$ as detectable SBBHs.
In our analysis, we set $\rho_{\mathrm{thr} }=1$, deviating from the commonly used thresholds of 8 or 12, as we focus on the accuracy requirement of different PN models to accurately detect as many SBBHs as possible.
Setting too high $\rho_{\mathrm{thr} }$ would result in a smaller number of detectable SBBHs, which would hinder our research objectives.

\begin{figure}[ht]
    \begin{minipage}{\columnwidth}
        \centering
        \includegraphics[width=0.9\textwidth,
        trim=0 0 0 0,clip]{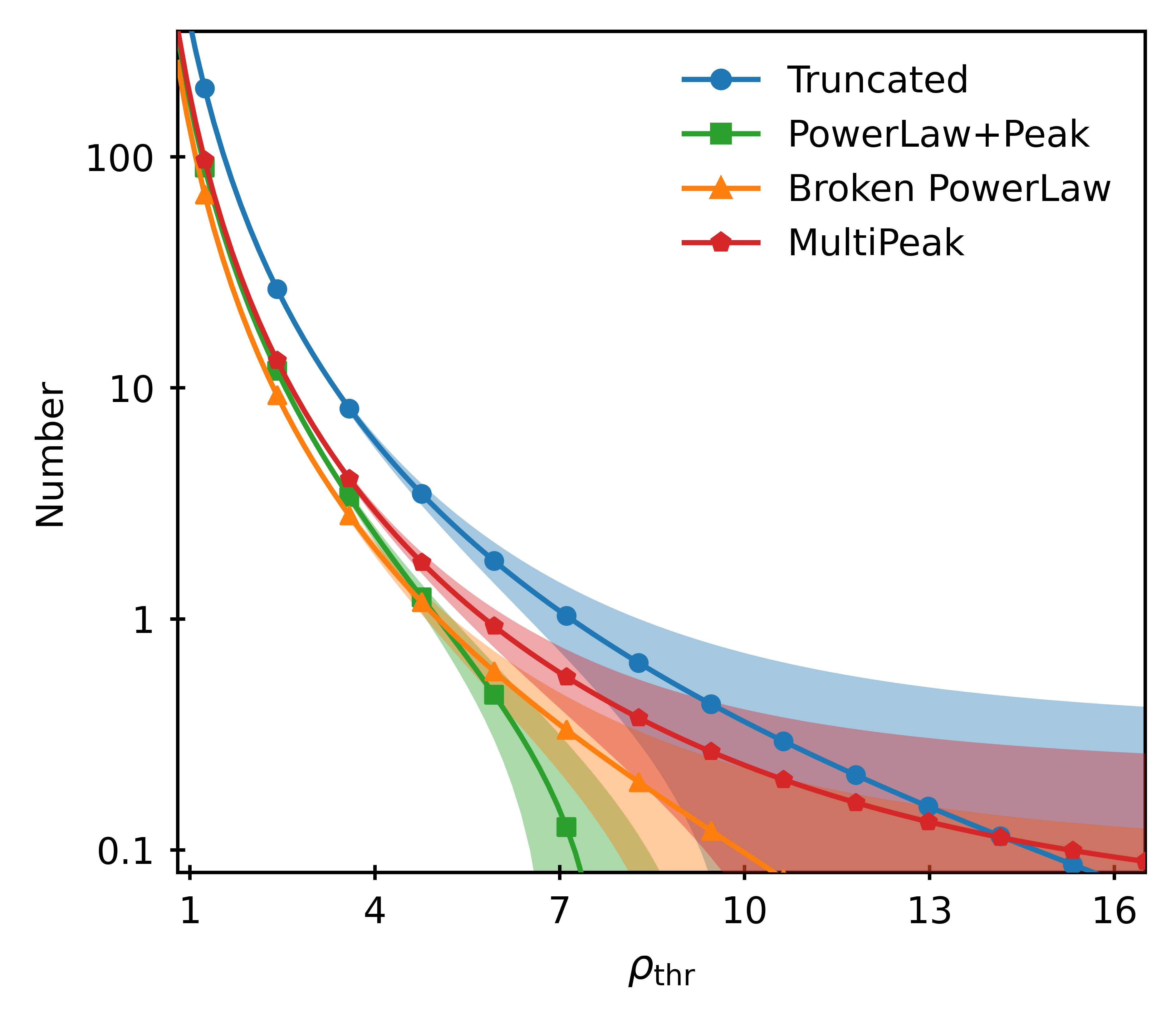}
        \caption{The number of detectable SBBHs by LISA under different SNR thresholds $\rho_{\mathrm{thr}}$ using four population models. The curve is the result of fitting using the equation $N_{\mathrm{thr} }=a\cdot \rho_{\mathrm{thr} }^{-3}+b$, and the shaded region represents the 90\% confidence interval of the fitting result.}\label{Fig:number}
    \end{minipage}
\end{figure}

\begin{figure}[ht]
    \begin{minipage}{\columnwidth}
        \centering
        \includegraphics[width=0.96\textwidth,
        trim=0 0 0 0,clip]{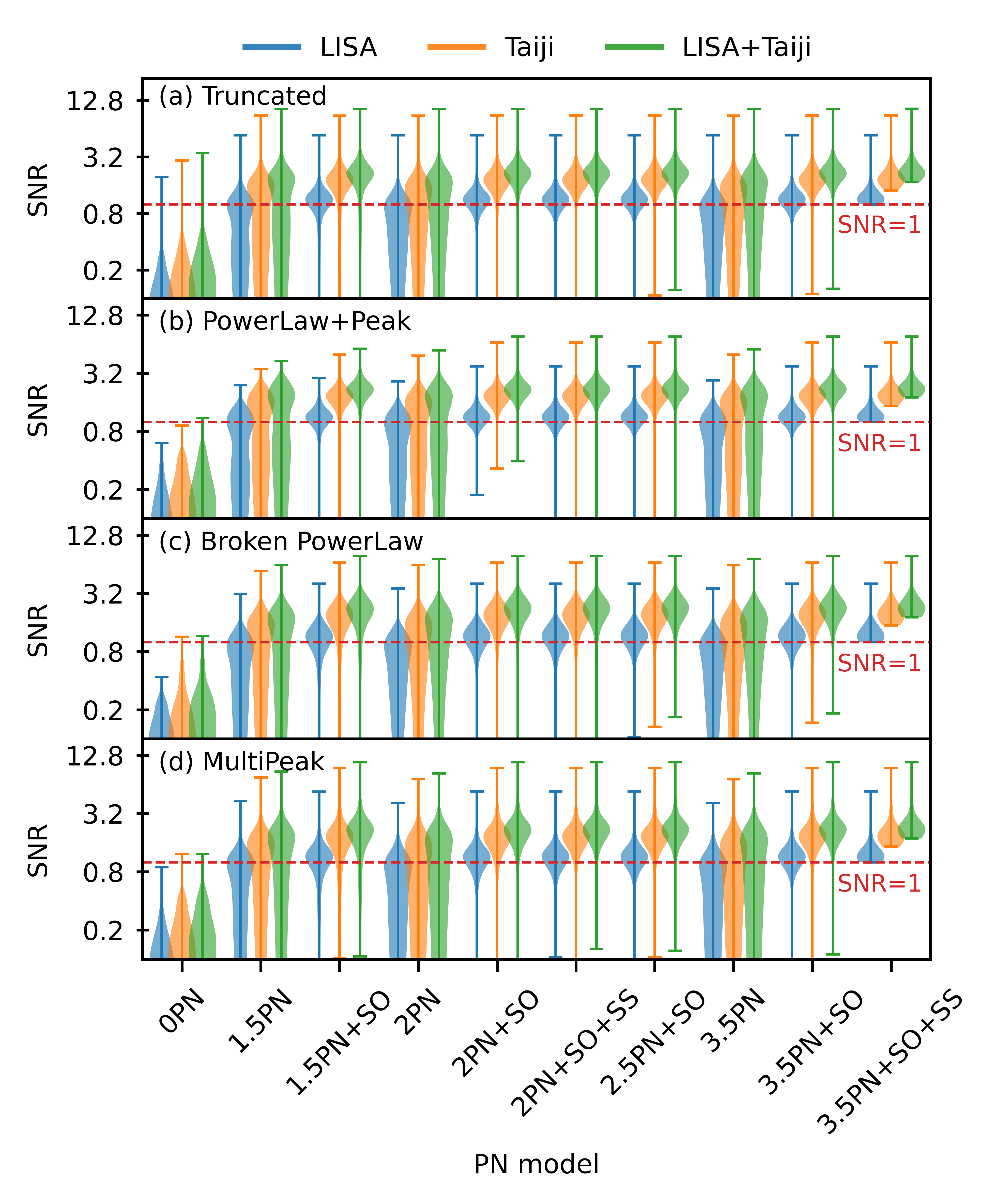}
        \caption{SNR distributions obtained from different PN models in the four population models. We present SNR distributions for three scenarios: LISA, Taiji, and LISA+Taiji. All considered SBBHs are detectable sources for LISA with $\rho\ge 1$.}\label{Fig:SNR}
    \end{minipage}
\end{figure}

Within the context of LISA, we consider SBBHs with $\rho_{\mathrm{opt} }\ge\rho_\mathrm{thr}$ to be detectable.
That standard is consistently applied throughout our study, including the examination of Taiji and joint detection in Sec.~\ref{Accuracy of different PN models}.
The relevant detailed research on detectable numbers can be referred to in Refs.~\cite{LISA_band,LISA_population,multiband1,multiband2,multiband3,TianQin_SBBH,LISA_Taiji}, and our results are shown in Fig.~\ref{Fig:number}.

According to Refs.~\cite{SNR_thr1,SNR_thr2}, the probability distribution function of SNR follows a power-law form, $p(\rho)\propto\rho^{-4}$. 
Therefore, the approximate representation of the detectable numbers $N_{\mathrm{thr} }$ above SNR threshold $\rho_{\mathrm{thr} }$ can be expressed as~\cite{LISA_population}
\begin{equation}\label{Eq:number}
    N_{\mathrm{thr}}\propto\int_{\rho>\rho_{\mathrm{thr}}}{\rho^{-4}}\propto\rho_{\mathrm{thr}}^{-3}.
\end{equation}
From Eq.~(\ref{Eq:number}), as the threshold increases, the number of detectable SBBHs decreases significantly, indicating the feasibility of choosing $\rho_{\mathrm{thr} }=1$.
Based on Fig.~\ref{Fig:number}, the detectable numbers show little variation among the four mass distribution models.
Additionally, Fig.~\ref{Fig:SNR} presents SNR distributions for different PN models, providing further insights into detection.

Fig.~\ref{Fig:SNR} shows that, for different PN models, except for 0PN model, the SNR of most models is higher than the threshold of $\rho_{\mathrm{thr} }=1$.
PN models with spin effects demonstrate a clear advantage over those without spin effects, resulting in more GW signals with SNR greater than $\rho_{\mathrm{thr} }$.
Additionally, under the same conditions, joint detection results are the best, while Taiji's results are better than LISA's, as detailed in Table.~\ref{Tab:detection_accuracy}.

\begin{figure*}[ht]
    \begin{minipage}{\textwidth}
        \centering
        \includegraphics[width=0.92\textwidth,
        trim=0 0 0 0,clip]{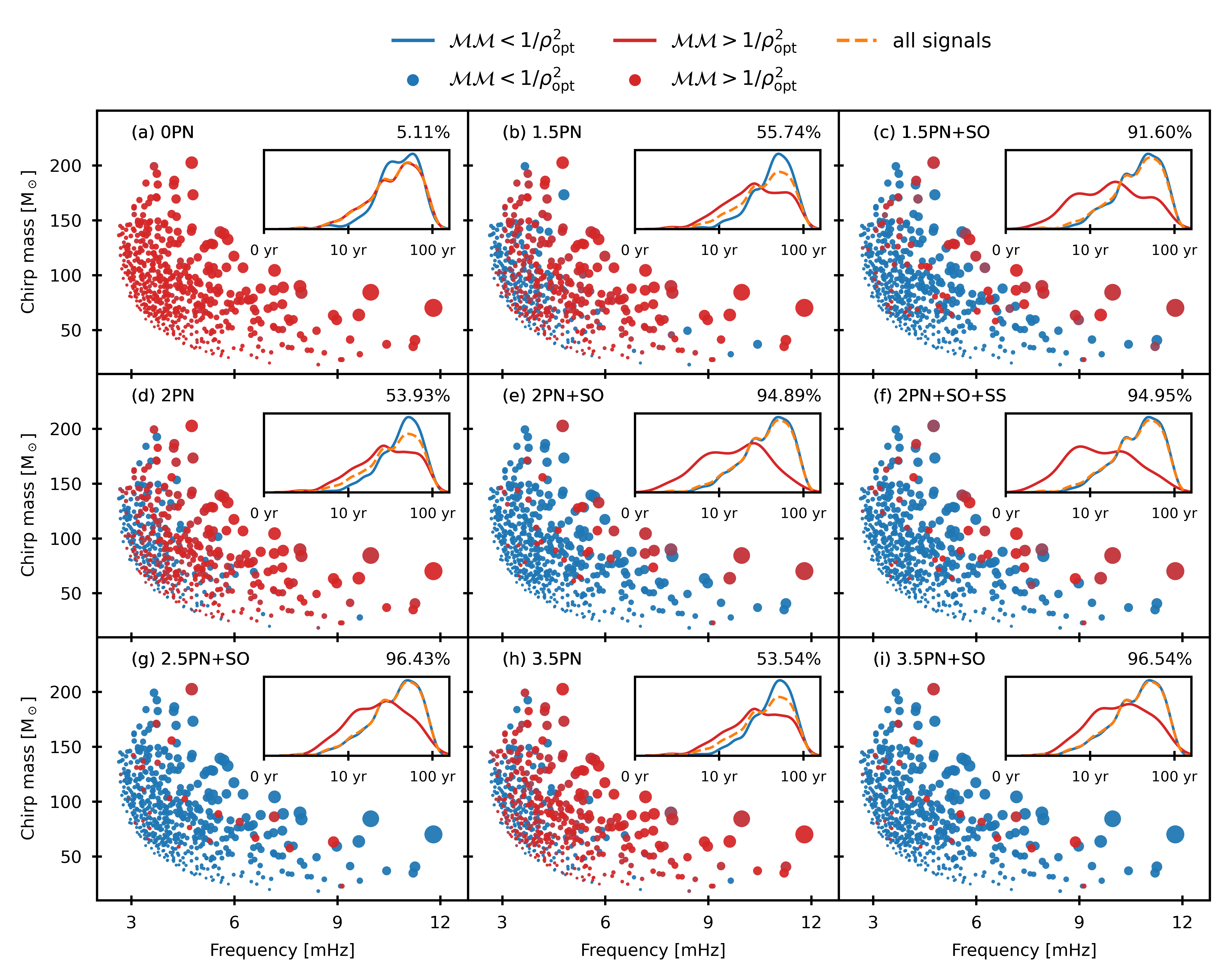}
        \caption{Comparison and distribution of accuracy requirement for different PN models. We provide the distribution of GW signals in the four population models with LISA, Taiji, and LISA+Taiji, using Eq.~(\ref{Eq:Accuracy1}) as the standard for accuracy requirement. The subplots uses blue to indicate compliance with accuracy requirements, and red to indicate non-compliance. The size of the points represents the value of $\tau$, where larger points indicate smaller values of $\tau$, and vice versa. The small chart in each subplot displays the normalized distribution of $\tau$, with accompanying numbers indicating the percentage of GW signals that meet accuracy requirement.}\label{Fig:Mc_f_tau_MM}
    \end{minipage}
\end{figure*}

\subsection{Accuracy of different PN models}\label{Accuracy of different PN models}
In Sec.~\ref{Detection number}, we generate datasets for four population models and extract GW signals with $\rho\ge1$.
Then we analyze the accuracy requirement of those GW signals with respect to different PN models according to Eq.~(\ref{Eq:Accuracy1}), and present them in Fig.~\ref{Fig:Mc_f_tau_MM}.
The data in Fig.~\ref{Fig:Mc_f_tau_MM} shows results similar to those in Fig.~\ref{Fig:SNR}.
PN models with spin effects yield more GW signals meeting accuracy requirement than those without spin effects.
Additionally, the proportion of GW signals that meet accurate requirement is greater in high-order PN models than in low-order ones.

For GW signals that do not meet accuracy requirement (red points), it is a challenge to establish a definitive and absolute threshold for determining the parameter values (e.g., $m_1$, $m_2$, $D_L$, etc.) whether meet accuracy requirement.
An insightful trend can be drawn from the distribution of $\tau$ shown in the small charts of Fig.~\ref{Fig:Mc_f_tau_MM}.
Specifically, GW signals that do not meet accuracy requirement tend to have smaller $\tau$ values than those meeting the requirement.
That is consistent with PN formalism, where larger $\tau$ values correspond to a longer duration until the binary system merges, resulting in a closer PN approximation to the true physical GW signal.
Certainly, that trend is statistical, and there can still be GW signals with small $\tau$ values meeting accuracy requirement.
More stringent accuracy requirement and detailed results considering different detectors can be found in Table.~\ref{Tab:detection_accuracy}.

\begin{table*}[ht]
\centering
\renewcommand{\arraystretch}{1.5}
\caption{The proportion of detection rate and accuracy requirements for different PN models with LISA, Taiji, and LISA+Taiji.}\label{Tab:detection_accuracy}
\begin{tabular*}{\textwidth}{@{\extracolsep{\fill}}cccccccccc@{}}
\hline
& \multicolumn{3}{c}{detection rate [\%]} & \multicolumn{3}{c}{$\mathcal{MM}< 1/\rho^2 _{\mathrm{opt} } $ [\%]} & \multicolumn{3}{c}{$\left \langle \delta h|\delta h \right \rangle <1$ [\%]} \\
PN model & ~~~LISA~~~ & ~~~Taiji~~ & LISA+Taiji & ~~~LISA~~~ & ~~~Taiji~~ & LISA+Taiji & ~~~LISA~~~ & ~~~Taiji~~ & LISA+Taiji\\
\hline
0PN
& 0.49    & 0.82     & \multicolumn{1}{c|}{1.48} 
& 15.32    & 0     & \multicolumn{1}{c|}{0} 
& 0    & 0     & 0          \\
1.5PN
& 38.88    & 67.38     & \multicolumn{1}{c|}{69.85} 
& 78.09    & 47.12     & \multicolumn{1}{c|}{42.01} 
& 54.53    & 34.93     & 29.49          \\
1.5PN+SO
& 87.48    & 92.09     & \multicolumn{1}{c|}{93.57} 
& 96.54    & 89.79     & \multicolumn{1}{c|}{88.47} 
& 90.61    & 87.81     & 87.48          \\
2PN
& 35.42    & 66.56     & \multicolumn{1}{c|}{70.51} 
& 78.42    & 44.98     & \multicolumn{1}{c|}{38.39} 
& 51.24    & 31.47     & 26.52          \\
2PN+SO
& 91.76    & 95.22     & \multicolumn{1}{c|}{96.38} 
& 97.69    & 93.57     & \multicolumn{1}{c|}{93.41} 
& 94.23    & 92.09     & 91.27          \\
2PN+SO+SS
& 91.60    & 95.22     & \multicolumn{1}{c|}{96.21} 
& 97.36    & 94.23     & \multicolumn{1}{c|}{93.25} 
& 94.40    & 92.75     & 91.93          \\
2.5PN+SO
& 94.07    & 95.88     & \multicolumn{1}{c|}{96.71} 
& 98.52    & 95.55     & \multicolumn{1}{c|}{95.22} 
& 95.72    & 94.73     & 94.23        \\
3.5PN
& 35.42    & 66.56     & \multicolumn{1}{c|}{71.0} 
& 78.42    & 44.81     & \multicolumn{1}{c|}{37.4} 
& 51.57    & 31.47     & 26.52        \\
3.5PN+SO
& 94.23    & 96.05     & \multicolumn{1}{c|}{96.87} 
& 98.52    & 95.72     & \multicolumn{1}{c|}{95.39} 
& 96.05    & 95.06    & 94.89          \\
\hline
\end{tabular*}
\end{table*}

In order to define detection rate, we calculate SNR $\rho$ for various PN models with Eq.~(\ref{Eq:SNR_overlap}) with a threshold $\rho_{\mathrm{thr} }=1$, and the proportion of detectable GW signals for different PN models relative to the total number of GW signals detectable by LISA.
We do not consider GW signals that can be detected through Taiji or joint detection, but cannot be detected through LISA.
According to Eqs.~(\ref{Eq:Accuracy1}) and (\ref{Eq:Accuracy2}), the proportions of GW signals that meet the different accuracy requirements can be calculated.
Overall, Table~\ref{Tab:detection_accuracy} illuminates the results about detection rate and two accuracy requirements of different PN models in LISA, Taiji and LISA+Taiji, which is consistent with the results shown in Figs.~\ref{Fig:SNR} and \ref{Fig:Mc_f_tau_MM}.
From Eqs.~(\ref{Eq:dis_noise_LISA}) and (\ref{Eq:dis_noise_Taiji}), Taiji has a better sensitivity than LISA for the lower displacement noise, therefore, in terms of detection rates, Taiji is expected to detect more GW signals than LISA.
Furthermore, joint detection using multiple detectors is expected to result in a higher detection rate than individual detectors.
For PN models with spin effects, except for 1.5PN+SO model in LISA with a detection rate of 87.5\%, the detection rates for the other PN models are consistently above 90\%.
The high-order PN model result in only a slight improvement in detection rates.

For accuracy requirement, a better sensitivity implies stricter constraints, and thus, the joint detection has the strictest constraints, while Taiji's constraints are stricter than LISA's. 
Additionally, the requirement of Eq.~(\ref{Eq:Accuracy2}) is stricter than that of Eq.~(\ref{Eq:Accuracy1}).
Similar to the detection rates, the differences in results for PN models with spin effects are not significant. 
Even for 1.5PN+SO, using joint detection and the requirement $\left \langle \delta h|\delta h \right \rangle <1$, 87.5\% of GW signals meet accuracy requirement.
However, under the same conditions, 3.5PN+SO model achieves nearly 95\% of GW signals meeting accuracy requirement.
Considering the time comparison in Fig.~\ref{Fig:time}, it depends on specific needs and requirement to achieve the balance between computational resources and the accuracy requirement of the final results.

\begin{figure}[ht]
    \begin{minipage}{\columnwidth}
        \centering
        \includegraphics[width=0.95\textwidth,
        trim=0 0 0 0,clip]{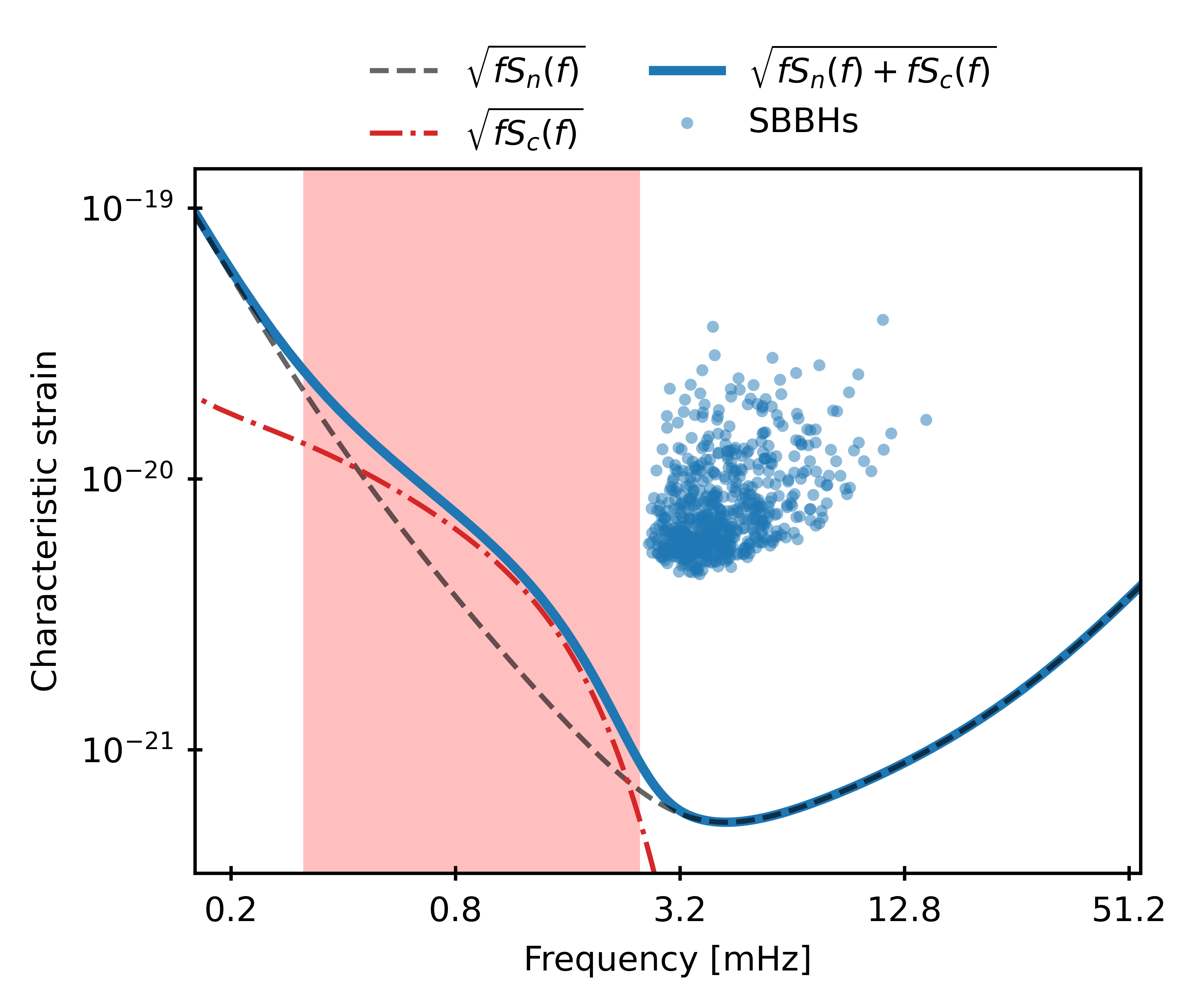}
        \caption{The impact of foreground noise on SBBHs. The dashed lines represent the foreground noise $S_c(f)$ and the PSD $S_n(f)$, while the solid line represents their combined result. We approximate the SBBHs with points, and the evolving tracks and non-evolving points of binary systems in the frequency spectrum are discussed in Ref.~\cite{LISA_noise2}. The red region represents the frequency band affected by foreground noise, whereas the green region represents the unaffected band.}\label{Fig:confusion_noise}
    \end{minipage}
\end{figure}

In the above calculations, we do not consider the impact of foreground noise $S_c(f)$ (also known as confusion noise or confusion foreground).
That is because based on our previous work~\cite{my_paper}, the foreground noise generated by galactic binaries (GBs) in the Milky Way falls within the frequency band of approximately $\sim$0.3-3~mHz, but the frequency distribution of the GW signals discussed in our paper shows little overlap with the foreground noise (see Figs.~\ref{Fig:Mc_f_tau_MM} and \ref{Fig:confusion_noise}).
Therefore, we do not separately consider the influence of foreground noise.
Furthermore, unlike the clearly defined spatial distribution (ecliptic coordinates) of GBs in the Milky Way, the spatial distribution of SBBHs assumes a random and uniform distribution, which means that different orbital configurations for Taiji have negligible differences.
Consequently, we only consider the typical orbital configuration of Taiji (i.e., Taiji-p)~\cite{Alternative_LISA-TAIJI,my_paper}.

\begin{figure*}[ht]
    \begin{minipage}{\textwidth}
        \centering
        \includegraphics[width=0.92\textwidth,
        trim=0 0 0 0,clip]{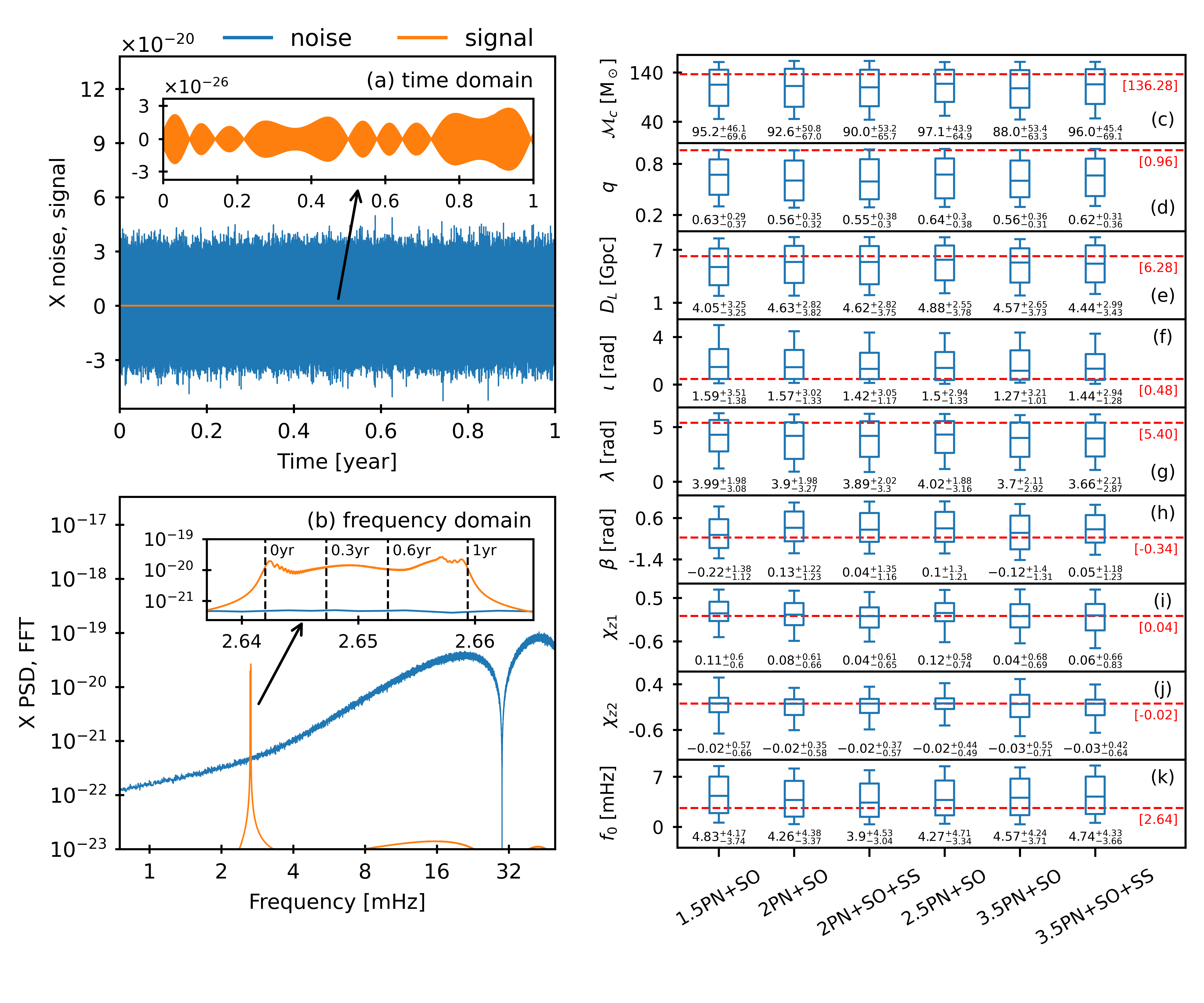}
        \caption{Parameter estimation of an GW signal using MCMC. \textit{Left panel}: Here are enhanced descriptions of the time domain and frequency domain images of GW signals. The yellow color represents the GW signal, while the blue color represents the instrumental noise using X-TDI 2.0. To provide a closer view, a small inset is used to provide a magnified view of a specific region. \textit{Right panel}: We estimate 9 parameters of GW source and visualize them using box plots. The upper and lower horizontal lines within each box plot represent the 90\% confidence interval. The edges of the box correspond to the upper and lower quartiles, while the line inside the box represents the median. Furthermore, the specific values for the median and the 90\% confidence interval are provided below each box. The red dashed line indicates the true parameters of GW source, with their exact values enclosed in square brackets on the right side.}\label{Fig:PE}
    \end{minipage}
\end{figure*}

\subsection{Parameter estimation}\label{Parameter estimation}
For parameter estimation, we randomly select a GW signal from the generated dataset with a SNR of approximately 1, which represents the most prevalent scenario.
The specific characteristics of GW signal are illustrated in Fig.~\ref{Fig:PE}.
Fig.~\ref{Fig:PE}(a) clearly shows that the GW signal in the time domain is overwhelmed by the instrumental noise.
The implementation of SNR above the detection threshold largely relies on long-term observations.
The inset in Fig.~\ref{Fig:PE}(b) reveals that the frequency variation of the GW signal is minimal, with a slight frequency variation of only 0.02~mHz over a period of one year.
Given those observations, it is reasonable to speculate that the differences among various PN models are unlikely to be obvious.

Using the method described in Sec.~\ref{Statistical inference}, we simulate data for the XYZ-TDI 2.0 channels and employ a prior range that allowed for fluctuations of approximately ±20\% around the actual parameters.
For parameter estimation, we utilize the \texttt{emcee} package~\cite{emcee} and set up 25 chains, with the likelihood function defined in Eq.~(\ref{Eq:likelihood}).
After running the MCMC algorithm\cite{PE_LISA} for several thousand steps, we obtain the results shown in Fig.~\ref{Fig:PE}.

Based on Fig.~\ref{Fig:PE}(c)-(k), the results align with our initial expectations.
There are no significant differences among the parameter estimates from different PN models, as they all fall within the 90\% confidence interval.
An interesting but unexpected result is that during the MCMC run, 3.5PN+SO+SS model, representing the exact physical GW signal injected, does not consistently yield better parameter estimation than other PN models.
For instance, in terms of $D_L$, the results from 2.5PN+SO model is closer to the true values.
Similarly, for $\beta$, 1.5PN+SO model, the lowest-order PN model used here surprisingly provides the closest estimate to the true value.
That result may be attributed to the extremely weak GW signal with low SNR, which can also be influenced by the selected population model and defined parameter ranges (e.g., $z$, $\tau_\mathrm{max} $, etc.), making few high SNR GW signals.

Furthermore, we also attempt MCMC analysis on GW signals with relatively higher SNR, while increasing the number of chains or steps. 
Same as previous results, the parameter estimation results from different PN models exhibited no significant differences.
Therefore, we have reason to believe that, in that particular MCMC parameter estimation, the choice of PN models does not significantly impact the final results.

\subsection{Eccentricity and spin-precession}\label{Eccentricity and precession}
All of our results above are based on the assumptions of zero eccentricity and non-precession, while in practice, the SBBH system has elliptical orbits and precession.
Depending on different formation mechanisms, the eccentricity and precession distribution of SBBH are different~\cite{Eccentricity_precession0}.
The SBBH formed through isolated binary evolution is expected to have a circular orbit, with its spin composition almost aligned with the binary angular momentum.
The dynamically formed SBBH in globular clusters (GC) or galactic nuclei may have non-zero eccentricity and random spin direction due to the influence of interactions in dense stellar environments~\cite{Eccentricity_precession1,Eccentricity_precession2}.

In the research on PN, the development of eccentric methods and spin-precessing methods has largely proceeded independently, and the development of models that include both eccentricity and precession is still in the initial stage~\cite{EFPE_ME}.
A recent study has announced an Effective-One-Body (EOB) Numerical-Relativity waveform model with eccentricity and precession~\cite{LiuXiaolin2}.
We employ the waveform code \texttt{pyWaveformGenerator}~\cite{LiuXiaolin1,LiuXiaolin2}, developed based on that EOB model, to investigate the influences of eccentricity and spin-precession.

\begin{table*}[ht]
\centering
\renewcommand{\arraystretch}{1.5}
\caption{Comparison of mismatches $\mathcal{MM}$ [\%] for GW waveforms with different eccentricity $e$ and spin parameters $\boldsymbol{\chi}_{1/2}$. We calculate the SBBH $\mathcal{MM}$ for different masses and mass ratios under each parameter with LISA. The values in the table represent the median of these results, and the upper and lower limits represent the confidence interval of 1 $\sigma$.}\label{Tab:eccentricity_precession}
\begin{tabular*}{\textwidth}{@{\extracolsep{\fill}}ccccccc@{}}
\hline
& \multicolumn{6}{c}{$\boldsymbol{\chi}_1:(\chi_{x1},\chi_{y1},\chi_{z1}),\boldsymbol{\chi}_2:(\chi_{x2},\chi_{y2},\chi_{z2})$} \\
$e$&$(0,0,0)$,$(0,0,0)$&$(0.1,0,0)$,$(0,0,0)$&$(0.1,0,0)$,$(0.1,0,0)$&$(0.1,0,0)$,$(-0.1,0,0)$&$(0.1,0,0)$,$(0,0.1,0)$&$(0.1,0,0)$,$(0.06,0.08,0)$\\ \hline
$0$&$0.0^{+0.0}_{-0.0}$&$0.91^{+1.97}_{-0.86}$&$3.15^{+1.78}_{-3.02}$&$0.51^{+6.98}_{-0.4}$&$1.08^{+3.88}_{-1.0}$&$2.48^{+4.84}_{-2.11}$\\
$10^{-3}$&$0.71^{+4.97}_{-0.68}$&$1.15^{+6.8}_{-1.05}$&$3.2^{+5.73}_{-3.12}$&$0.81^{+2.36}_{-0.76}$&$1.8^{+8.3}_{-1.76}$&$2.9^{+8.76}_{-2.76}$\\
$10^{-2}$&$2.05^{+8.22}_{-1.98}$&$6.67^{+10.06}_{-4.78}$&$11.93^{+6.25}_{-8.61}$&$7.83^{+11.18}_{-6.26}$&$13.73^{+9.79}_{-9.67}$&$11.37^{+8.78}_{-8.22}$\\
$0.1$&$11.35^{+5.24}_{-4.7}$&$13.16^{+10.48}_{-8.46}$&$12.75^{+8.53}_{-6.36}$&$13.68^{+10.88}_{-7.02}$&$12.08^{+9.16}_{-7.24}$&$12.77^{+6.98}_{-5.71}$\\ \hline
\end{tabular*}
\end{table*}

In the most sensitive frequency range of LISA, we use \texttt{pyWaveformGenerator} to calculate the mismatches under different eccentricity and spin parameters using the same method as in Sec.~\ref{Mismatch}, as shown in Table~\ref{Tab:eccentricity_precession}.
As expected, for eccentricity, a larger eccentricity will result in a larger $\mathcal{MM}$.
That is due to the GW generated by elliptical orbits, whose frequency is different from the monotonic variation of circular orbits and fluctuates within a certain range.
In addition, the larger the eccentricity, the greater the GW frequency fluctuation. 
According to Eq.~(\ref{Eq:wave_amp}), the amplitude also fluctuates, resulting in a larger $\mathcal{MM}$.
For spin-precession, different spin directions can lead to different results.
The influence of spin components in the same direction ($\chi_{x1}=\chi_{x2}=0.1$, $\chi_{y1}=\chi_{y2}=0$) on GW waveform is greater than that in different directions ($\chi_{x1}=\chi_{y2}=0.1$, $\chi_{y1}=\chi_{x2}=0$), with the opposite direction ($\chi_{x1}=-\chi_{x2}=0.1$, $\chi_{y1}=\chi_{y2}=0$) having the smallest impact.
This is similar to the spin effect in Eq.~(\ref{Eq:phase}), and there should be some degree of cancellation in different directions.
Moreover, the effect of spin-precession on GW frequency is not as significant as eccentricity, so the resulting $\mathcal{MM}$ is smaller than that of eccentricity.

For small impacts, $\mathcal{MM}$ can still be controlled within an acceptable range, but for large impacts, $\mathcal{MM}$ may reach a larger level, and the differences in GW waveform cannot be ignored.
However, when considering population models, our PN model without eccentricity and spin-precession is not useless.
As shown in Fig.~\ref{Fig:population}(c) and (d), the spin amplitudes in most SBBHs are very small, and the direction is likely to be biased towards the z-axis, resulting in very small x and y components that generate precession.
In Ref.~\cite{Eccentricity_precession1}, Kremer~\textit{et al} expect that within the frequency band of LISA, approximately 30\% of GC binary systems have an eccentricity exceeding $10^{-3}$, and 18\% have an eccentricity exceeding $10^{-2}$.
These results suggest that, besides isolated SBBHs with zero eccentricity and non-precession, most dynamically formed SBBHs also have extremely small eccentricity and non-aligned spin components.
Therefore, our PN model can still be used in the vast majority of SBBHs.

Due to the limitations of theoretical GW waveform development, we have achieved a relatively complete level at the current stage.
One reason is the enormous computational time required, so we can only conduct basic analysis and research on $\mathcal{MM}$ in Table.~\ref{Tab:eccentricity_precession}.
In spite of its limitations, the study certainly adds to our understanding of the differences in GW waveforms.
Further studies need to be carried out in order to comprehensively validate the effects of eccentricity and spin-precession on different PN models.

\section{Conclusion}\label{Conclusion}
In this paper, we investigate the comparison and application of different PN models in detecting SBBHs using space-based GW detectors.
Specifically, we consider PN models with spin effects and assume non-precessing system with circular orbits.
We examine three detector scenarios: LISA, Taiji, and their joint detection, considering the most general response functions and the Doppler effect caused by the motion around the Sun.
For the population model, we utilize the four SBBH population models provided by LVK and simulate detectable GW signals based on LISA's SNR threshold.
In terms of comparison, we conduct a comprehensive analysis of the different PN models based on three key aspects: mismatch, detection rate, and accuracy requirement.
Furthermore, we employ the MCMC method with TDI 2.0 to estimate the parameters of a GW signal, providing insights into the application of those models for parameter estimation.

Our research indicates that for PN models with spin effects, over 99\% of the cases achieved a high level of match with a Match of 0.994 or higher.
Moreover, a high Match is observed for SBBHs with small total mass, indicating superior performance of PN models for lower-mass SBBHs.
Given the prevalence of low mass ($\le 50\ \mathrm{M}_\odot$) in the population, 1.5PN+SO model is deemed sufficient for simulating population-level GW signals.

In terms of detection rate, LISA achieves a detection rate exceeding 90\% with 2PN+SO model, while Taiji achieves the same detection rate with 1.5PN+SO model, demonstrating the feasibility of employing lower-order PN models for GW detection.
Applying the most stringent accuracy requirement, we find that with 1.5PN+SO model, 90\% of GW signals meet accuracy requirement for LISA.
It implies that for 90\% of detectable GW signals, LISA can not distinguish between 1.5PN+SO model and the exact physical GW signal.
Even in the joint detection, 2PN+SO model still meets accuracy requirement for 91\% of GW signals.
From a statistical perspective, GW signals that meet accuracy requirement usually have larger $\tau $ values than those do not meet it.
In addition, 1.5PN+SO model achieves those results in half the time compared to 3.5PN model, while delivering superior outcomes.

Furthermore, in parameter estimation using MCMC Bayesian inference, there is no significant differences in the results obtained from different PN models.
Importantly, a short generation times for GW waveforms in high-dimensional parameter spaces implies a reduction of computing resources and running time.
In summary, the selection of an appropriate PN model hinges on achieving a balance between computational resources and the desired accuracy.

In our future research, we plan to make improvements in several aspects to enhance the study of SBBHs using space-based GW detectors. 
Firstly, we aim to further investigate the application of low-order PN models, which will help reduce computational resources.
In our previous paper~\cite{my_paper_LIGO}, we utilized a modified 0PN model for simulating data and actual GW signals in LIGO.
We plan to employ a similar approach for space-based GW detectors.
Secondly, compared to ground-based GW detectors, the influence of eccentricity on GW signals is more pronounced in space-based GW detectors~\cite{PN_all,eccentricity}.
Eccentric orbits introduce phase correction different from circular orbits, and we anticipate the need for higher-order PN models to accurately simulate this effect~\cite{PN_accuracy}.
In addition, the PN series exhibits some unpredictable alternating behavior, and recent works have pushed the waveform to 4.5PN~\cite{4_5PN1,4_5PN2,4_5PN3}.
Therefore, we can further extend our research to higher-order cases.
For cases beyond GR, the PN model may exhibit different effects.
For example, in the Einstein-Cartan theory, the correction of 1PN is considered detectable through the pulsar timing array technique, and we can study the situation in space-based GW detectors~\cite{Einstein-Cartan1,Einstein-Cartan2,Einstein-Cartan3,Einstein-Cartan4}.
Additionally, with the operation of LVK O4 Run, we expect to obtain more reliable astrophysical population models and better constraints on various parameters, which will contribute to further research on detectable SBBHs.
Finally, in MCMC Bayesian inference, we plan to explore different MCMC sampler or parameter estimation methods, such as \texttt{PyMC3}~\cite{PyMC3} and heterodyned likelihood method~\cite{Heterodyned}, to study the differences in application of different PN models.
In conclusion, with the aforementioned in-depth investigations, we believe we can advance the study of SBBHs, providing more valuable information for space-based GW detector observations and data analysis.

\begin{acknowledgements}
This work was supported by the National Key Research and Development Program of China (Grant No.~2021YFC2203004), the National Natural Science Foundation of China (Grant No.~12347101) and the Natural Science Foundation of Chongqing (Grant No.~CSTB2023NSCQ-MSX0103). 
\end{acknowledgements}

\bibliography{references}
\end{document}